\documentclass{JHEP3}
 \usepackage{amsthm,graphicx,amsfonts,amsmath,amssymb,latexsym,cancel}

 \theoremstyle{plain}
 \newtheorem {hypo}{\bf\hspace{-\parindent}Hypothesis}

 \newtheorem {lem}[hypo]{Lemma}
 \newtheorem {claim}[hypo]{Claim}
 \theoremstyle{definition}

 \newcommand{\pf}{\begin{bpf}}

 \newcommand{\pfms}{\begin{bpfms}}
 \newcommand{\epf}{\end{bpf}\hfill$\square$\vspace{0.1cm}}
 \newcommand{\epfms}{\end{bpfms}\hfill$\square$\\ }
 \newcommand\ben{\begin{equation*}}
 \newcommand\ebn{\end{equation*}}
 \newcommand\be{\begin{equation}}
 \newcommand\eb{\end{equation}}
 \newcommand\Zb{\mathbb{Z}}
 \newcommand\Rb{\mathbb{R}}
 \newcommand\Cb{\mathbb{C}}
 \newcommand\Pb{\mathbb{P}}

 \title{Painlev\'e VI connection problem and monodromy  of $c=1$ conformal blocks}

 \author{N. Iorgov$^{1}$, O. Lisovyy$^{1,2}$ and Yu. Tykhyy$^{1}$ \\
 \!\!\!$^1$\! Bogolyubov Institute for Theoretical Physics, 03680, Kyiv, Ukraine \\
 \!\!\!$^2$\! Laboratoire de Math\'ematiques et Physique Th\'eorique CNRS/UMR 7350,
 \\ Universit\'e de Tours, 37200 Tours, France
 \vspace{2mm} \\
 {\footnotesize \tt iorgov@bitp.kiev.ua, lisovyi@lmpt.univ-tours.fr, tykhyy@bitp.kiev.ua}}

 \abstract{Generic $c=1$ four-point conformal blocks on the Riemann sphere can be seen as the
 coefficients of Fourier expansion of the tau function
 of Painlev\'e VI equation with respect to one of its integration constants. Based on this relation,
 we show that $c=1$ fusion matrix essentially coincides with
 the connection coefficient relating tau function asymptotics at different critical points.
 Explicit formulas for both quantities are obtained by solving certain functional relations which follow from the
 tau function expansions. The final result does not involve integration and
 is given by a ratio of two products of Barnes $G$-functions with arguments
 expressed in terms of conformal dimensions/monodromy data. It turns out to be closely related to the volume
 of hyperbolic tetrahedron.
 }

 \makeatletter\let\default@color\current@color\makeatother

 \keywords{Conformal field theory, Painlev\'e equations}
 \preprint{}

\begin{document}

 \section{Introduction}
 The two-dimensional conformal field theory (CFT) \cite{BPZ} has been intensively
 studied in the last three decades. A renewed interest to these
 studies is related to the recent discovery \cite{AGT} of a relation between 2D CFTs
 and $\mathcal{N}=2$ 4D supersymmetric gauge theories, commonly referred to as AGT correspondence.

 The infinite-dimensional conformal symmetry determines the structure
 of correlation functions and leads to the notion of conformal blocks: these are universal chiral parts of correlators
 corresponding to different choices of intermediate conformal families in the  successive operator product expansions (OPE) of primary fields.
 From a mathematical standpoint,
 conformal blocks can be seen as new special functions arising in the representation theory
 of the Virasoro algebra. The AGT relation provides us with their explicit series representations.

 Equivalence of different ways to decompose a correlation function into a sum over conformal families suggests the existence
 of duality transformations of conformal blocks, formalized by the concept of Moore-Seiberg groupoid \cite{MS}. In particular,
 there should exist an elementary invertible linear map connecting $s$- and $t$-channel four-point Virasoro conformal blocks
 on the Riemann sphere, AGT-related to weak/strong coupling $S$-duality on the gauge side.
 The integral kernel of this transformation is called the fusion matrix. Its explicit form
 was obtained in \cite{pt1,pt2} by solving certain functional equations (which follow from the
 Moore-Seiberg formalism) with the help of representation theory of the
 modular double of $\mathcal{U}_q(\mathfrak{sl}(2,\mathbb{R}))$. An alternative derivation, based on
 free-field representations of chiral vertex operators, was proposed later in
 \cite{TLiou,Tlecture}.

 The results of \cite{pt1,pt2,TLiou,Tlecture} hold for generic complex values of the Virasoro central charge $c$.
 Unfortunately, it is not clear whether/how  they can be extended to the half-line $c\in\mathbb{R}_{\leq 1}$, including a particularly interesting point
 $c=1$ \cite{RW,Schomerus,ER} at the borderline between minimal models and Liouville theory.

 The present work approaches the last problem by exploiting the relation of $c=1$ conformal
 blocks and Painlev\'e VI equation \cite{CFT_PVI,p35}. It turns out to be mutually useful.
 We will show that $c=1$ fusion matrix essentially coincides with
 a connection coefficient for Painle\-v\'e~VI tau function expressed in terms of monodromy data of the auxiliary linear problem.
 Conformal expansions of the tau function imply that this coefficient satisfies certain recurrence relations. On the other hand, equivalence
 of different critical points of Painlev\'e VI can be seen as a kind of crossing symmetry condition.
 Connecting expansion parameters in different channels, it makes the recurrence relations highly nontrivial and restrictive.
 Their solution appears to be related to the Poisson geometry of the moduli space
 of monodromy data and complexified volume
 of generic hyperbolic tetrahedron.

 It is worth mentioning that the connection problem for tau functions of Painlev\'e equations has a strong independent interest.
 Such questions arise, e.g. in the study of the large gap asymptotics of Fredholm determinants of integrable kernels arising in random matrix theory
 \cite{baik,dikz,dyson,dyson2f1}.  In this framework, the analogs of the connection coefficients are called Dyson constants.
 Their computation involves integrals of the classical Painlev\'e transcendents and so far seemed to be inaccessible
 with the existing tools of Painlev\'e theory. Most of the available exact results have been obtained on case by case basis
 by approximating the corresponding Fredholm determinants with Toeplitz and Hankel determinants \cite{basor,cik,dik,ehrhardt,krasovsky}.
 Hopefully, our results will provide some new insight in this context.

 The paper is organized as follows. In Section~\ref{seccb}, we recall basic symmetry properties of conformal blocks, explain
 Ponsot-Teschner formula for the fusion kernel for generic $c$ and discuss a few explicit examples. In Section~\ref{secpvi}, after a brief
 outline of
 the relation between $c=1$ conformal blocks and Painlev\'e VI, we discuss monodromy data for the associated linear problem
 and their relation to hyperbolic tetrahedron. Connection problem for Painlev\'e VI tau function is solved in Section~\ref{seccc}. Its main result is the explicit formula (\ref{concofans}) for the connection coefficient.
 The latter is related to $c=1$ fusion matrix in Section~\ref{secfusion}, see formula (\ref{fusionfinal}).
 The proofs of some technical results are relegated to Appendix.

 \acknowledgments{We are grateful to O. Gamayun, P. Gavrylenko, A. Morozov, A. Mironov, S. Ribault and J. Teschner for helpful discussions and comments
 on the manuscript. The present work  was supported by the Ukrainian SFFR projects F53.2/028 (N.I.) and F54.1/019 (Yu.T.), the Program of fundamental research of
 the physics and astronomy division of NASU, and the IRSES project ``Random and integrable models in mathematical physics'' (O.L.).}
 \section{Conformal blocks}\label{seccb}
 \subsection{Symmetries}
 Let us start by fixing some notation. Throughout this paper, we use a Liouville-type parameterization
 of the central charge:
 \ben
 c=1+6Q^2,\qquad Q=b+b^{-1}.
 \ebn
 To cover all possible complex values of $c$, it suffices to consider $b$ from the first quadrant. The weak-coupling region $c\geq 25$ then corresponds to $b\in\mathbb{R}_{\geq1}$, the values $c\leq 1$ to $b\in i\mathbb{R}_{\geq1}$, and
  $1\leq c\leq 25$ to a quarter of the unit circle $b=e^{i\varphi}$, $\varphi\in[0,\frac{\pi}{2}]$.
 It is convenient to represent conformal weights of primary fields in the form
 \ben
 \Delta=\frac{c-1}{24}+\theta^2,
 \ebn
 where the parameters $\theta$ will be referred to as momenta.

 Four-point $s$-channel Virasoro conformal block on $\Pb^1$ with external dimensions  $\Delta_{\nu}=\frac{c-1}{24}+\theta_{\nu}^2$
 attached to the points $\nu=0,t,1,\infty$ and internal dimension $\Delta_{\sigma}=\frac{c-1}{24}+\sigma^2$ will be
 written in one of the following forms:
 \ben
 \mathcal{F}_{c} \left(\{\Delta_{\nu}\},\Delta_{\sigma};t\right)
 =\mathcal{F}_{c}\Bigl[\begin{array}{ll}\theta_1 & \theta_t \\ \theta_{\infty} & \theta_0\end{array};\sigma\Bigr]\left(t\right)=
 \vcenter{\hbox{\includegraphics[height=10ex]{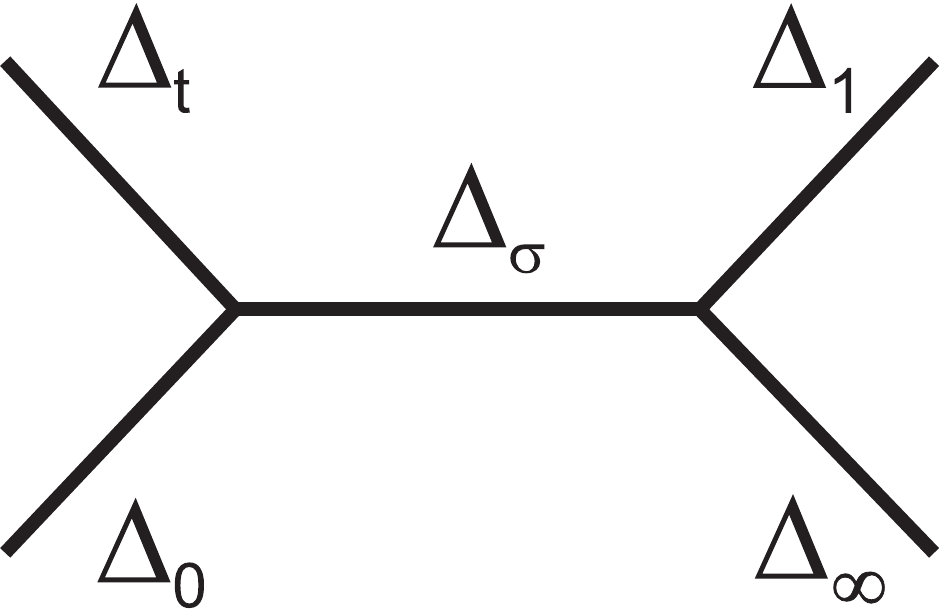}}}
 \ebn
 It is a power series in $t$ normalized as $\mathcal{F}_c\left(\{\Delta_{\nu}\},\Delta_{\sigma};0\right)=1$.

 As a function of $t$, $\mathcal{F}_c$ is believed to be analytically continuable to the universal cover of $\Pb^1\backslash\{0,1,\infty\}$. Some intuition about this analytic behavior
 may be gained by looking at the limit $c\rightarrow\infty$, $\Delta$'s finite,
 where conformal block reduces to Gauss hypergeometric series
 \ben
 \mathcal{F}_{\infty}\left(\{\Delta_{\nu}\},\Delta_{\sigma};t\right)={}_2F_1\left(\Delta_t-\Delta_0+\Delta_{\sigma},
 \Delta_1-\Delta_{\infty}+\Delta_{\sigma};2\Delta_{\sigma};t\right).
 \ebn
 The present paper mainly deals with another special case $c=1$, where conformal block function
 becomes a Fourier transform of the tau function of the sixth Painlev\'e equation
 with respect to
 one of its integration constants.

 As a function of parameters, conformal block enjoys a number of symmetries, analogous to Euler-Pfaff
 fractional linear transformations of $_2F_1\left(a,b;c;z\right)$:
 \begin{itemize}
 \item \textit{Trivial symmetries}. Changing the sign of any of $\theta_{0,t,1,\infty}$ or $\sigma$
 has no effect on conformal block as the latter depends on dimensions only.
 \item \textit{$R$-symmetries} allow the exchange of columns and rows of external mo\-menta:
 \begin{align}
 \label{Rsym1}
 &\mathcal{F}_{c}\Bigl[\begin{array}{ll}\theta_1 & \theta_t \\ \theta_{\infty} & \theta_0\end{array};\sigma\Bigr]\left(t\right)=\left(1-t\right)^{\Delta_0-\Delta_t-\Delta_1+\Delta_{\infty}}
 \mathcal{F}_{c}\Bigl[\begin{array}{ll}\theta_{\infty} & \theta_0 \\ \theta_{1} & \theta_{t}\end{array};\sigma\Bigr]\left(t\right)=\\
 \label{Rsym2}
 =\,&\mathcal{F}_{c}\Bigl[\begin{array}{ll}\theta_t & \theta_1 \\ \theta_{0} & \theta_{\infty}\end{array};\sigma\Bigr]\left(t\right)=\left(1-t\right)^{\Delta_0-\Delta_t-\Delta_1+\Delta_{\infty}}
 \mathcal{F}_{c}\Bigl[\begin{array}{ll}\theta_{0} & \theta_{\infty} \\ \theta_{t} & \theta_{1}\end{array};\sigma\Bigr]\left(t\right).
 \end{align}
 \item \textit{$T$-symmetry} enables one to exchange the dimensions in one column:
 \be\label{Tsym}
 \mathcal{F}_{c}\Bigl[\begin{array}{ll}\theta_1 & \theta_t \\ \theta_{\infty} & \theta_0\end{array};\sigma\Bigr]\left(t\right)=\left(1-t\right)^{\Delta_0-\Delta_t-\Delta_{\sigma}}
 \mathcal{F}_{c}\Bigl[\begin{array}{ll}\theta_{\infty} & \theta_t \\ \theta_{1} & \theta_0\end{array};\sigma\Bigr]\left(\frac{t}{t-1}\right).
 \eb
 \item \textit{Regge-Okamoto symmetry}. There is an identity
 \be\label{regge}
 \mathcal{F}_{c}\Bigl[\begin{array}{ll}\,\theta_1-\delta & \theta_t-\delta \\ \theta_{\infty}-\delta & \theta_0-\delta\end{array};\sigma\Bigr]\left(t\right)=\left(1-t\right)^{\delta_{1t}\delta}
 \mathcal{F}_{c}\Bigl[\begin{array}{ll}\theta_1 & \theta_t \\
 \theta_{\infty} & \theta_0\end{array};\sigma\Bigr]\left(t\right),
 \eb
 where $\delta$, $\delta_{1t}$ are defined by
 \ben
  2\delta=\theta_0+\theta_t+\theta_1+\theta_{\infty},\qquad \delta_{1t}=\theta_t+\theta_1-\theta_0-\theta_{\infty}.
 \ebn
 This is reminiscent of the unexpected Regge symmetry of Racah-Wigner $6j$ symbols and Okamoto symmetry of Painlev\'e~VI \cite{boalchR}. The latter can be actually seen as a $c=1$ specialization of the above.
 Though it is not easy to derive (\ref{regge}) from CFT first principles, this relation becomes almost obvious
 in the AGT representation where it corresponds to a permutation of masses of matter hypermultiplets.
 Being combined with trivial symmetries, it relates conformal blocks with three distinct (unordered)
 sets of external dimensions.
 \end{itemize}

 \subsection{Linear transformations}
 Conformal blocks appear in the expansion of the four-point correlator of primary fields with additional prefactors. It is convenient to introduce the function
 \be\label{extcb}
 \bar{\mathcal{F}}_{c}\Bigl[\begin{array}{ll}\theta_1 & \theta_t \\ \theta_{\infty} & \theta_0\end{array};\sigma\Bigr]\left(t\right)=t^{\Delta_{\sigma}-\Delta_0-\Delta_t}
 \mathcal{F}_{c}\Bigl[\begin{array}{ll}\theta_1 & \theta_t \\ \theta_{\infty} & \theta_0\end{array};\sigma\Bigr]\left(t\right),
 \eb
 defined on $\Pb^1\backslash\{(-\infty,0]\cup[1,\infty)\}$ with the choice of the principal branch for the fractional
 powers of $t$.

 It is useful to think of the variable $t$ as being the cross-ratio $\displaystyle t=\frac{\left(z_2-z_1\right)\left(z_4-z_3\right)}{
 \left(z_3-z_1\right)\left(z_4-z_2\right)}$ of four points $z_1=0$, $z_2=t$, $z_3=1$, $z_4=\infty$. The mapping class group
 $\Gamma=PSL_2\left(\Zb\right)$
 of the 4-punctured sphere is the quotient of the braid group on 3 strings by its center. It naturally acts on conformal blocks by braiding transformations of $z_{1,2,3,4}$ and
 appropriate permutations of dimensions. One of the ge\-ne\-rators of this action is
 given by the above $T$-transformation. The second
 generator acts as
 $S:\; \theta_0\leftrightarrow\theta_1,\;t\leftrightarrow 1-t$. It can be checked that $S$ and $T$ satisfy the modular group defining relations $S^2=(ST)^3=1$.

 It is expected that the linear span of conformal blocks (\ref{extcb}) with different internal dimensions realizes
 an infinite-dimensional representation of $\Gamma$ due to associativity of the operator product expansions.
 More precisely, there should be a linear
 ``$S$-duality'' relation between the conformal
 blocks calculated in different channels:
 \be\label{fusion}
 \bar{\mathcal{F}}_{c}\Bigl[\begin{array}{ll}\theta_1 & \theta_t \\ \theta_{\infty} & \theta_0\end{array};\sigma\Bigr]\left(t\right)=\int_{\Rb^+}
 F_{c}\Bigl[\begin{array}{ll}\theta_1 & \theta_t \\ \theta_{\infty} & \theta_0\end{array};\begin{array}{c}\rho \\ \sigma\end{array}\Bigr]
 \bar{\mathcal{F}}_{c}\Bigl[\begin{array}{ll}\theta_0 & \theta_t \\ \theta_{\infty} & \theta_1\end{array};\rho\Bigr]\left(1-t\right)\,d\rho.
 \eb
 The $t$-independent kernel $F_c$  is the  fusion matrix. It may be assumed to be even function
 of parameters $\theta_{\nu}$, $\sigma$, $\rho$ and has a number of symmetries similar to
 (\ref{Rsym1}), (\ref{Rsym2}) and (\ref{regge}).

 The explicit form of the fusion kernel was found by Ponsot and Teschner
 who identified it with the Racah-Wigner matrix for a class
 of infinite-dimensional representations of the quantum group $\mathcal{U}_q\left(\mathfrak{sl}\left(2,\mathbb{R}\right)\right)$
 \cite{pt1,pt2}. Their result reads
 \begin{align}
 \nonumber
 F_{c}\Bigl[\begin{array}{ll}\theta_1 & \theta_t \\ \theta_{\infty} & \theta_0\end{array};\begin{array}{c}\rho \\ \sigma\end{array}\Bigr]=
 \prod_{\epsilon,\epsilon'=\pm}\hat{\Gamma}_b\biggl[\begin{array}{c}{
 \epsilon{\theta}_1-\theta}_t+\epsilon'{\rho},\epsilon{\theta}_0+{\theta}_{\infty}+\epsilon'{\rho}\\
  \epsilon{\theta}_0-{\theta}_t+\epsilon'{\sigma},\epsilon{\theta}_1+{\theta}_{\infty}+\epsilon'{\sigma}
 \end{array}\biggr]
 \prod_{\epsilon=\pm}\hat{\Gamma}_b\biggl[\begin{array}{c}2\epsilon{\sigma}-\frac{iQ}{2}\\ 2\epsilon{\rho}+\frac{iQ}{2}\end{array}\biggr]&\,\times\\
 \label{ptf}
 \times\int_{\mathcal{C}}dx\;\prod_{\epsilon=\pm}
  \hat{S}_b\biggl[\begin{array}{c} \frac{iQ}{2}+\epsilon{\theta}_0-{\theta}_t+x,\frac{iQ}{2}+\epsilon{\theta}_1+{\theta}_{\infty}+x\\
 \epsilon{\sigma}+x,{\theta}_{\infty}-{\theta}_{t}+\epsilon{\rho}+x\end{array}\biggr]&\,.
 \end{align}
 where we use the standard convention $f\Bigl[\begin{array}{c}\alpha_1,\ldots,\alpha_n \\ \beta_1,\ldots,\beta_m\end{array}\Bigr]=\displaystyle \frac{\prod_{k=1}^n f\left(\alpha_k\right)}{\prod_{k=1}^m f\left(\beta_k\right)}$.

 The functions
 $\hat{\Gamma}_b(x)$ and
 $\hat{S}_b(x)=\hat{\Gamma}_b\left(x\right)/\hat{\Gamma}_b\left(-x\right)$ are closely related to
 the Barnes double gamma function and quantum dilogarithm. They can be defined by analytic continuation of
 the integral representations
 \begin{align*}
 \ln\hat{\Gamma}_b\left(x\right)&\,=\int_0^{\infty}\frac{dt}{t}\left\{\frac{e^{-2ixt}-1}{4\sinh bt\sinh b^{-1}t}+\frac12 x^2 e^{-2t}
 +\frac{ix}{2t}\right\},\\
 \ln\hat{S}_b\left(x\right)&\,=\int_0^{\infty}\frac{dt}{it}\left\{\frac{\sin2xt}{2\sinh bt\sinh b^{-1}t}
 -\frac{x}{t}\right\}.
 \end{align*}
 The function $\hat{S}_b(x)$ has an infinite number of zeros and poles in the complex $x$-plane:
 \begin{itemize}
 \item zeros: $x=-ib\left(m+\frac12\right)-ib^{-1}\left(n+\frac12\right)$ with $m,n\in\Zb_{\geq0}$,
 \item poles: $x=ib\left(m+\frac12\right)+ib^{-1}\left(n+\frac12\right)$ with $m,n\in\Zb_{\geq0}$.
 \end{itemize}
 This implies that, for instance, for real $b\geq1$ the integrand in (\ref{ptf}) has eight infinite half-lines of poles
 shown in Fig. 1a. As $b=e^{i\varphi}$, $\varphi\in\left(0,\frac{\pi}{2}\right)$, the half-lines open to 2D lattice sectors, see
 Fig. 1b. Similar picture holds for any $b$ with $\mathrm{Re}\,b>0$.

 The integration contour $\mathcal{C}$ in (\ref{ptf}) runs from $-\infty$ to $+\infty$ passing between
 the upper and lower pole sectors. With this prescription, Ponsot-Teschner formula gives the fusion kernel for any complex value of the
 central charge \textbf{except} for the half-line $c\in\Rb_{\leq 1}$ corresponding to purely imaginary $b$. The present paper is mainly concerned with
 the edge point $c=1$ of this excluded region.

     \begin{figure}[!h]
 \begin{center}
 \resizebox{14cm}{!}{
 \includegraphics{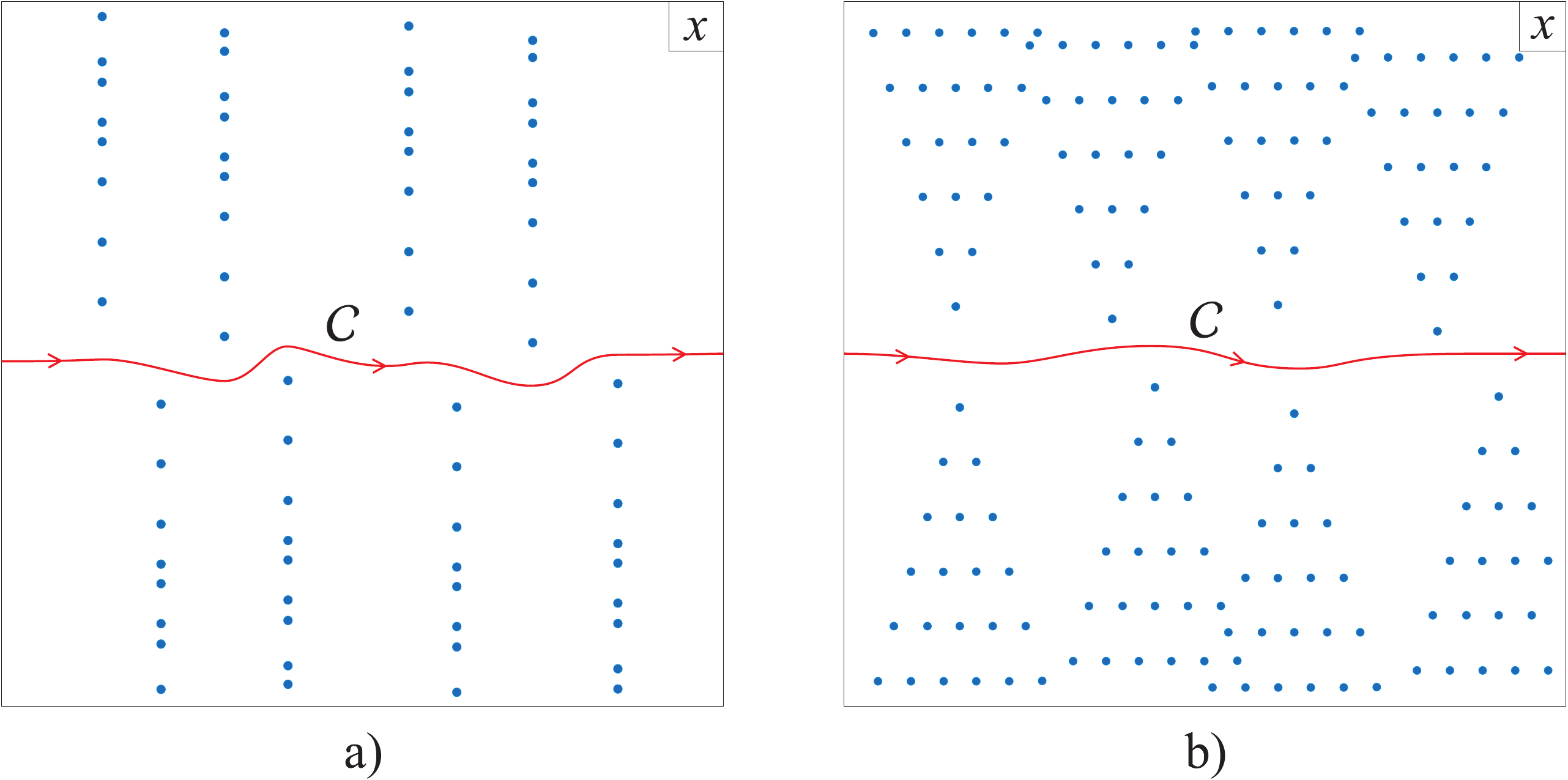}} \\
 {\small Fig. 1: Integrand singularities and integration contour in Ponsot-Teschner formula \\ for a) $c>25$ and b) $1<c<25$ }
 \end{center}
 \end{figure}

 \subsection{Checking Ponsot-Teschner formula: an  example with $c=25$}\label{ssc25}
 Let us illustrate the formula (\ref{ptf}) for the fusion kernel with an explicit example. It will be based on
 the evaluation of conformal block with $c=25$, arbitrary internal dimension and
 all external dimensions equal to $\frac{15}{16}$, found by Al.~B.~Zamolodchikov \cite[footnote (1)]{Zamo_AT}.
 This corresponds to setting $b=1$ and $\theta_{0,t,1,\infty}=\frac{i}{4}$ in the above.

 The answer appears in \cite{Zamo_AT} in a parameterization particularly suitable for the modular transformations, which may be explained as follows. Consider a complex torus $\Cb/\left(\mathbb{Z}+\tau\Zb\right)$ and identify its points related by multiplication by $-1$.
 This yields a double cover of $\Pb^1$ with 4 ramification points. Their cross-ratio (our variable $t$) remains invariant under the
 action of the subgroup $\Gamma(2)$ corresponding to pure analytic continuation of conformal blocks. It is explicitly given by
 the elliptic lambda function
 \be\label{tintermsoftau}
 t=\frac{\vartheta_2^4\left(0|\tau\right)}{\vartheta_3^4\left(0|\tau\right)},
 \eb
 where $\vartheta_{2,3}\left(z|\tau\right)$ are the usual Jacobi theta functions
 \begin{align*}
 \vartheta_2\left(z|\tau\right)=\sum_{n\in\Zb+\frac12}e^{i\pi n^2\tau+2inz},\qquad
 \vartheta_3\left(z|\tau\right)=\sum_{n\in\Zb}e^{i\pi n^2\tau+2inz}.
 \end{align*}
 The inverse map can be written as
 \be\label{tauintermsoft}
 \tau=i\frac{K(1-t)}{K(t)},
 \eb
 where $K(t)$ denotes complete elliptic integral of the 1st kind:
 \ben
 K(t)=\frac{\pi}{2}{}_2F_1\left(\frac12,\frac12;1;t\right)=\int_0^1\frac{dx}{\sqrt{(1-x^2)(1-tx^2)}}.
 \ebn
 It is clear that in the limit $t\rightarrow +0$ one has $\tau\rightarrow i\infty$. Even more specifically, one
 can check that $\lim\limits_{t\rightarrow+0}t^{-1}e^{i\pi\tau}=\frac{1}{16}$.

 Now, using (\ref{tauintermsoft}), the aforementioned result of \cite{Zamo_AT}  can be stated as
 \be\label{c25cb}
 \bar{\mathcal{F}}_{25}\Bigl[\begin{array}{ll}\frac{i}{4} & \frac{i}{4} \\ \frac{i}{4} & \frac{i}{4}\end{array};\sigma\Bigr]\left(t\right)=\frac{2^{4\sigma^2}e^{i\pi\sigma^2\tau}}{t^{\frac78}\left(1-t\right)^{\frac78}
 \vartheta_3^3\left(0|\tau\right)}.
 \eb
 The modular transformation $S$ exchanging $t$ and $1-t$ maps $\tau$ to $-\tau^{-1}$. Then, applying Jacobi's imaginary transformation
 to (\ref{c25cb}),
 it is straightforward to verify that the $S$-duality relation (\ref{fusion}) is satisfied by
 Fourier transform conjugated by simple diagonal factors
 \be\label{fusion25}
 F_{25}\Bigl[\begin{array}{cc}\frac{i}{4} & \frac{i}{4} \\ \frac{i}{4} & \frac{i}{4}\end{array};
 \begin{array}{c}\rho \\ \sigma\end{array}\Bigr]=\left(2^{-4\sigma^2}\sigma\right)^{-1}2\sin2\pi\sigma\rho\;\left(2^{-4\rho^2}\rho\right).
 \eb

 Next let us try to derive this relation from the
 Ponsot-Teschner formula. The functions $\hat{\Gamma}_b(x)$ and $\hat{S}_b(x)$ in
 the limit $b\rightarrow 1$ are expressed by means of the Barnes $G$-function:
 \begin{align*}
 \hat{\Gamma}_{b\rightarrow1}(x)&\,=\frac{\left(2\pi\right)^{\frac{ix}{2}}}{G(1+ix)},\\
 \hat{S}_{b\rightarrow1}(x)&\,=\left(2\pi\right)^{ix}\frac{G(1-ix)}{G(1+ix)}.
 \end{align*}
 Thanks to the doubling identity
 \ben
 G\left(1+2x\right)=2^{x(2x-1)}\pi^{-x-\frac12}G\biggl[
 \begin{array}{c}
 \frac12+x,1+x,1+x,\frac32+x \\ \frac12,\frac12
 \end{array}\biggr]
 \ebn
 the prefactor in the 1st line of (\ref{ptf}) reduces to
 \ben
 \left(2^{-4\sigma^2}\sigma\right)^{-1}
 4\sinh2\pi\sigma\sinh2\pi\rho\;\left(2^{-4\rho^2}\rho\right).
 \ebn
 Similarly simplifying the integrand in the 2nd line, it is possible to show that the fusion kernel (\ref{fusion25}) will
 follow from (\ref{ptf}) provided that
  \be\label{straf}
 \int_{\mathcal{C}}\frac{dx}{\left(2\pi\right)^{2ix}}\frac{G(1-2ix)}{G(1+2ix)}
 \prod_{\epsilon=\pm}G\biggl[
 \begin{array}{cc}
 2+i\epsilon \sigma+ix, & 2+i\epsilon\rho+ix \\
 i\epsilon\sigma-ix, & i\epsilon\rho-ix
 \end{array}
 \biggr]=\frac{8\pi^4\sin2\pi\sigma\rho}{\sinh2\pi\sigma\sinh2\pi\rho}.
 \eb
 The contour $\mathcal{C}$ passes between the half-lines of zeros of the Barnes functions in the denominator,
 as discussed above.
 In particular, for $\sigma,\rho\in\Rb$ it can be chosen as horizontal line with $0<\mathrm{Im}\,x<\frac12$.

 The intriguing integral identity (\ref{straf}) looks rather nontrivial and we will not attempt to rigorously prove it here.
 Instead, we contented ourselves with its numerical verification for several randomly chosen values of $\sigma$ and $\rho$.

 \subsection{Probing $c=1$ fusion}\label{subsectionAT}
 As $c$ approaches the interval $(-\infty,1]$, Ponsot-Teschner formula becomes singular.
 The sectors in Fig.~1b transform into overlapping half-planes containing an infinite number of dense lines of poles. The function $\hat{\Gamma}_b(x)$ has a
 natural boundary of analyticity at purely imaginary $b$.  It is therefore legitimate to ask whether fusion transformations merely exist. Of course, they do  for degenerate dimensions \cite{FD}. The present work suggests that this is also true for arbitrary
 dimensions at $c=1$.

 As an illustration, consider the Ashkin-Teller conformal block, characterized by $c=1$ and all external dimensions
 equal to $\frac{1}{16}$. This is the second solvable case where conformal block function
 is known in a closed form for arbitrary
  internal dimension \cite[Eq. (2.28)]{Zamo_AT}. In
 the elliptic parameterization (\ref{tintermsoftau})--(\ref{tauintermsoft}), one has
  \be\label{c1cb}
 \bar{\mathcal{F}}_{1}\Bigl[\begin{array}{ll}\frac{1}{4} & \frac{1}{4} \\ \frac{1}{4} & \frac{1}{4}\end{array};\sigma\Bigr]\left(t\right)=
 \frac{2^{4\sigma^2}e^{i\pi\sigma^2\tau}}{t^{\frac18}\left(1-t\right)^{\frac18}
 \vartheta_3\left(0|\tau\right)}.
 \eb
 It is very easy to check by evaluating Gaussian integrals that (\ref{fusion}) is again satisfied by a Fourier-type  fusion kernel
  \be\label{fusionAT}
 F_{1}\Bigl[\begin{array}{cc}\frac{1}{4} & \frac{1}{4} \\ \frac{1}{4} & \frac{1}{4}\end{array};
 \begin{array}{c}\rho \\ \sigma\end{array}\Bigr]=2^{4\sigma^2-4\rho^2+1} \cos2\pi\sigma\rho.
 \eb

 \section{Painlev\'e VI and hyperbolic tetrahedron}\label{secpvi}
 \subsection{Conformal expansions}
 The problem of determining the $c=1$ fusion matrix will
 be reformulated in Subsection~\ref{connectiontofusion} as connection problem for the tau function $\tau(t)$ of the sixth Painlev\'e equation
  \begin{align}
 \label{zetapvi}
 \nonumber\Bigl(t(t-1)\zeta''\Bigr)^2=-2\;\mathrm{det}\left(\begin{array}{ccc}
 2\theta_0^2 & t\zeta'-\zeta & \zeta'+\theta_0^2+\theta_t^2+\theta_1^2-\theta_{\infty}^2 \\
  t\zeta'-\zeta & 2 \theta_t^2 & (t-1)\zeta'-\zeta \\
  \zeta'+\theta_0^2+\theta_t^2+\theta_1^2-\theta_{\infty}^2 & (t-1)\zeta'-\zeta & 2\theta_1^2
 \end{array}\right),
 \end{align}
 defined by its logarithmic derivative
 \be\label{tauzeta}
 \displaystyle\zeta(t)=t\left(t-1\right)\frac{d}{dt}\ln\tau(t).
 \eb

 The relation of $\tau(t)$ to generic four-point $c=1$ conformal blocks was observed in \cite{CFT_PVI,p35}.
 Painlev\'e VI parameters $\vec{\theta}=\left(\theta_0,\theta_t,\theta_1,\theta_{\infty}\right)$ correspond to the external
 momenta, one of the constants of integration encodes the intermediate dimension spectrum and the
 other one is a generating parameter. Specifically, the tau function can be written~as
 \begin{align}
 \label{exp0}
 \tau(t)&=\chi_0\sum_{n\in\Zb}
 C\Bigl[\begin{array}{ll}\theta_1 & \theta_t \\ \theta_{\infty} & \theta_0\end{array};\sigma_{0t}+n\Bigr]\,
 s^n_{0t}\,
 \bar{\mathcal{F}}_{1}\Bigl[\begin{array}{ll}\theta_1 & \theta_t \\ \theta_{\infty} & \theta_0\end{array};\sigma_{0t}+n\Bigr]\left(t\right)=\\
 \label{exp1}
 &=\chi_1\sum_{n\in\Zb}
 C\Bigl[\begin{array}{ll}\theta_0 & \theta_t \\ \theta_{\infty} & \theta_1\end{array};\sigma_{1t}+n\Bigr]\,
 {s}^n_{1t}\,
 \bar{\mathcal{F}}_{1}\Bigl[\begin{array}{ll}\theta_0 & \theta_t \\ \theta_{\infty} & \theta_1\end{array};\sigma_{1t}+n\Bigr]\left(1-t\right),
 \end{align}
 The first representation is particularly suitable in the vicinity of $t=0$, and the second one gives
 the expansion of $\tau(t)$ around $t=1$. The structure constants are given by
 \be\label{cconst}
 C\Bigl[\begin{array}{ll}\theta_1 & \theta_t \\ \theta_{\infty} & \theta_0\end{array};\sigma\Bigr]=
 \frac{\prod_{\epsilon,\epsilon'=\pm}G(1+\theta_t+\epsilon\theta_0+\epsilon'\sigma)
 \,G(1+\theta_1+\epsilon\theta_{\infty}+\epsilon'\sigma)}{\prod_{\epsilon=\pm}G(1+2\epsilon\sigma)}.
 \eb
 Any of the two pairs of integration constants $\left(\sigma_{0t},s_{0t}\right)$ and $\left(\sigma_{1t},s_{1t}\right)$ specifies
 the initial conditions for Painlev\'e~VI in the form of solution asymptotics near a given critical point. The relation between the two pairs is most conveniently formulated in
 terms of monodromy data for the associated rank~2 linear problem which we shall now briefly discuss.

 \subsection{Monodromy and initial conditions}
 The space of monodromy data consists of conjugacy classes of triples $\left(\mathcal{M}_0,\mathcal{M}_t,\mathcal{M}_1\right)$ of monodromy matrices from $SL(2,\Cb)$.
 To describe it efficiently, one
 needs to introduce in addition to $\sigma_{0t}$, $\sigma_{1t}$ a third exponent $\sigma_{01}$ which appears in the expansion at $\infty$.
 These exponents and parameters $\vec{\theta}$ are related to monodromy matrices as follows:
 \begin{align*}
 \begin{array}{ll}
 p_{\mu}\;\, =2\cos2\pi\theta_{\mu}=\mathrm{Tr}\,\mathcal{M}_{\mu},& \qquad\qquad  \mu\;\,=0,t,1,\infty,\\
 p_{\mu\nu}=2\cos2\pi\sigma_{\mu\nu}=\mathrm{Tr}\,\mathcal{M}_{\mu}\mathcal{M}_{\nu}, &\qquad\qquad \mu\nu=0t,1t,01,
 \end{array}
 \end{align*}
 with $\mathcal{M}_{\infty}=\left(\mathcal{M}_{1}\mathcal{M}_{t}\mathcal{M}_{0}\right)^{-1}$.

 We define Painlev\'e VI monodromy manifold $\mathcal{M}$  as the corresponding $SL(2,\Cb)$-character variety of
 $\pi_1\left(\Pb^1\backslash\{0,t,1,\infty\}\right)$.
 It is described by the Jimbo-Fricke affine cubic surface
 $ W(p_{0t},p_{1t},p_{01})=0$,
 where \cite{jimbo}
 \begin{align}\label{JFr}
  W(p_{0t},p_{1t},p_{01})=p_{0t}p_{1t}p_{01}+p_{0t}^2+p_{1t}^2+p_{01}^2-
 \omega_{0t} p_{0t}-\omega_{1t} p_{1t}-\omega_{01} p_{01}+\omega_4-4.
 \end{align}
 The parameters $\vec{\omega}=\left(\omega_{0t},\omega_{1t},\omega_{01},\omega_4\right)$ depend
 only on $\vec{\theta}$ appearing in Painlev\'e~VI and are considered as fixed. They are explicitly given by
 \begin{align*}
 \omega_{0t}&\,=p_0p_t+p_1p_{\infty},\\ \omega_{1t}&\,=p_tp_1+p_0p_{\infty},\\ \omega_{01}&\,= p_0p_1+p_tp_{\infty},\\
 \omega_4&\,=\!\!\!\prod_{\mu=0,t,1,\infty}\!\!\!p_{\mu}+\!\!\!\sum_{\mu=0,t,1,\infty}\!\!\!p_{\mu}^2.
 \end{align*}

 The triples $\vec{\sigma}=\left(\sigma_{0t},\sigma_{1t},\sigma_{01}\right)$ satisfying the constraint $W(p_{0t},p_{1t},p_{01})=0$ pa\-ra\-mete\-rize the two-dimensional space of Painlev\'e~VI initial conditions. Fixing $p_{0t}$ in this constraint gives a quadric which admits rational parameterization. The quantity $s_{0t}$ in (\ref{exp0}) can be seen as the corresponding uniformizing parameter. The quantity $s_{1t}$ from (\ref{exp1}) plays a similar role if one fixes $p_{1t}$ instead of $p_{0t}$.
 Specifically,  $s_{0t}$ and $s_{1t}$ have the following expression in terms of monodromy \cite{jimbo}\footnote{After appropriate corrections: see
 Remark~25 in \cite{klein} and Remark~6 in \cite{dyson2f1}.}:
  \begin{align}
 \label{s0t}
 s_{0t}^{\pm1}=\frac{q_{01}e^{\pm 2\pi i\sigma_{0t}}-q_{1t}}{
 16\prod\limits_{\epsilon=\pm}
 \sin\pi\left(\theta_t\mp\sigma_{0t}+\epsilon\theta_0\right)
 \sin\pi\left(\theta_1\mp\sigma_{0t}+\epsilon\theta_{\infty}\right)},\\
 \label{s1t}
 s_{1t}^{\pm1}=\frac{q_{01}e^{\mp2\pi i\sigma_{1t}}-q_{0t}}{16\prod\limits_{\epsilon=\pm}
 \sin\pi\left(\theta_t\mp\sigma_{1t}+\epsilon\theta_1\right)\sin\pi\left(\theta_0\mp\sigma_{1t}+\epsilon\theta_{\infty}\right)},
 \end{align}
 where
 we have introduced the notation $\displaystyle q_{\mu\nu}=\frac{\partial W}{\partial p_{\mu\nu}}$
 so that, for instance,
 \be\label{q01def}
 q_{01}=2p_{01}+p_{0t}p_{1t}-\omega_{01}.
 \eb
 It turns out that Jimbo-Fricke cubic may be rewritten in terms of these variables in a nice determi\-nan\-tal form, e.g.,
 \be\label{JFr2}
 q_{01}^2=\mathrm{det}\,\mathcal{G},\qquad \mathcal{G}=\left(\begin{array}{cccc}
 2 & -p_0 & -p_t & p_{1t} \\
 -p_0 & 2 & p_{0t} & -p_{\infty} \\
 -p_t & p_{0t} & 2 & -p_1 \\
 p_{1t} & -p_{\infty} & -p_1 & 2\end{array}\right).
 \eb
 Appendix contains several useful relations involving first
 minors of the matrix  (\ref{JFr2}).
  In particular, they ensure consistency of the different sign choices in (\ref{s0t}), (\ref{s1t}).

 \subsection{Connection problem}
 The definition (\ref{tauzeta}) of the Painlev\'e~VI tau function contains an obvious normalization ambiguity, which
 implies that the coefficients $\chi_{0,1}$ in (\ref{exp0})--(\ref{exp1}) are intrinsically indefinite. However, their \textit{ratio}
 \be\label{concof}
 \chi_{01}(\vec{\theta};\sigma_{0t},\sigma_{1t};p_{01})=\chi_0^{-1}\chi_1
 \eb
 is completely fixed by the differential equation and initial conditions for $\zeta(t)$.
 It determines \textit{relative} normalization
 of the expansions of $\tau(t)$ near $0$ and $1$, and will be called connection coefficient.

 Sometimes it becomes convenient to include the structure constants into the definition
 of relative normalization by introducing
 \begin{align}
  \label{concof2}
 \bar{\chi}_{01}(\vec{\theta};\sigma_{0t},\sigma_{1t};p_{01})=\chi_{01}(\vec{\theta};\sigma_{0t},\sigma_{1t};p_{01})\,
 C\Bigl[\begin{array}{ll}\theta_0 & \theta_t \\ \theta_{\infty} & \theta_1\end{array};\sigma_{1t}\Bigr]\Bigl/
 C\Bigl[\begin{array}{ll}\theta_1 & \theta_t \\ \theta_{\infty} & \theta_0\end{array};\sigma_{0t}\Bigr].
 \end{align}
 In particular,
 for $-\frac12<\mathrm{Re}\,\sigma_{0t},\,\mathrm{Re}\,\sigma_{1t}<\frac12$ one can write
 \begin{align}
  \label{concof3}
 \bar{\chi}_{01}(\vec{\theta};\sigma_{0t},\sigma_{1t};p_{01})=\frac{
 \lim_{t\rightarrow 1}(1-t)^{\theta_1^2+\theta_t^2-\sigma_{1t}^2}\tau(t)}{
 \lim_{t\rightarrow 0}t^{\theta_0^2+\theta_t^2-\sigma_{0t}^2}\tau(t)
 }.
 \end{align}
  Finding explicit form of the connection coefficients (\ref{concof}), (\ref{concof2}) in terms of monodromy data
 constitutes one of the main goals of the present work.

 \subsection{Canonical coordinates}\label{seccancoords}
 There is a natural Poisson bracket $\{,\}$ on monodromy manifold $\mathcal{M}$ inherited from the Atiyah-Bott
 symplectic structure on the moduli space of flat logarithmic $SL(2,\Cb)$-connections on the 4-punctured sphere.
 This bracket is defined~by
 \ben
 \left\{p_{0t},p_{1t}\right\}=q_{01},\qquad \left\{p_{1t},p_{01}\right\}=q_{0t},\qquad
 \left\{p_{01},p_{0t}\right\}=q_{1t},
 \ebn
 with $q_{\mu\nu}$ the same as above. Parameterizing $s_{0t}$, $s_{1t}$ from (\ref{s0t}), (\ref{s1t}) as
 \be\label{etas}
 s_{0t}=e^{\eta_{0t}/2\pi i},\qquad s_{1t}=e^{\eta_{1t} /2\pi i},
 \eb
 it can be easily verified (we have learned  this from a recent work \cite{NRS}
 containing an equivalent observation) that the local coordinates $\eta_{0t}$, $\eta_{1t}$ are conjugate to
 monodromy exponents $\sigma_{0t}$,~$\sigma_{1t}$:
 \ben
 \left\{\sigma_{0t},\eta_{0t}\right\}=\{\sigma_{1t},\eta_{1t}\}=1.
 \ebn

 Two pairs of Darboux coordinates $\left(\sigma_{0t},\eta_{0t}\right)$
 and $\left(\sigma_{1t},\eta_{1t}\right)$ are well-adapted for characterizing the expansions of $\tau(t)$
 near $t=0$ and $t=1$, respectively, cf (\ref{exp0})--(\ref{exp1}).
 Observe that generic $c=1$ four-point Virasoro conformal blocks, as functions of $t$ and $1-t$, literally coincide with Fourier expansion coefficients of the appropriately normalized Painlev\'e~VI tau functions with respect to the dual coordinates $\eta_{0t}$ and $\eta_{1t}$.

 The pairs $\left(\sigma_{0t},\eta_{0t}\right)$
 and $\left(\sigma_{1t},\eta_{1t}\right)$ are related by a canonical transformation whose generating function
 $\mathcal{S}(\vec{\theta};\sigma_{0t},\sigma_{1t})$ is
 defined by the equations
 \be\label{genefun}
 \eta_{0t}=\frac{\partial \mathcal{S}}{\partial \sigma_{0t}},\qquad \eta_{1t}=-\frac{\partial \mathcal{S}}{\partial{\sigma_{1t}}}.
 \eb
 Remarkably, $\mathcal{S}(\vec{\theta};\sigma_{0t},\sigma_{1t})$ can be found in explicit form.
 It essentially coincides \cite{NRS} with the complexified volume of the hyperbolic tetrahedron $\mathcal{T}$
 with dihedral angles $2\pi\vec{\theta}$, $2\pi\sigma_{0t}+\pi$, $2\pi\sigma_{1t}+\pi$, whose
 mnemonic graphical representation can be obtained by gluing external legs of $s$- or $t$-channel conformal blocks (see Fig.~2).
 A tetrahedral signature shows up already in (\ref{JFr2}):  the $4\times 4$ matrix $\mathcal{G}$ is nothing but the Gram
 matrix of scalar products of length $\sqrt{2}$ vectors  normal to faces of $\mathcal{T}$ and oriented outwards.

    \begin{figure}[!h]
 \begin{center}
 \resizebox{4cm}{!}{
 \includegraphics{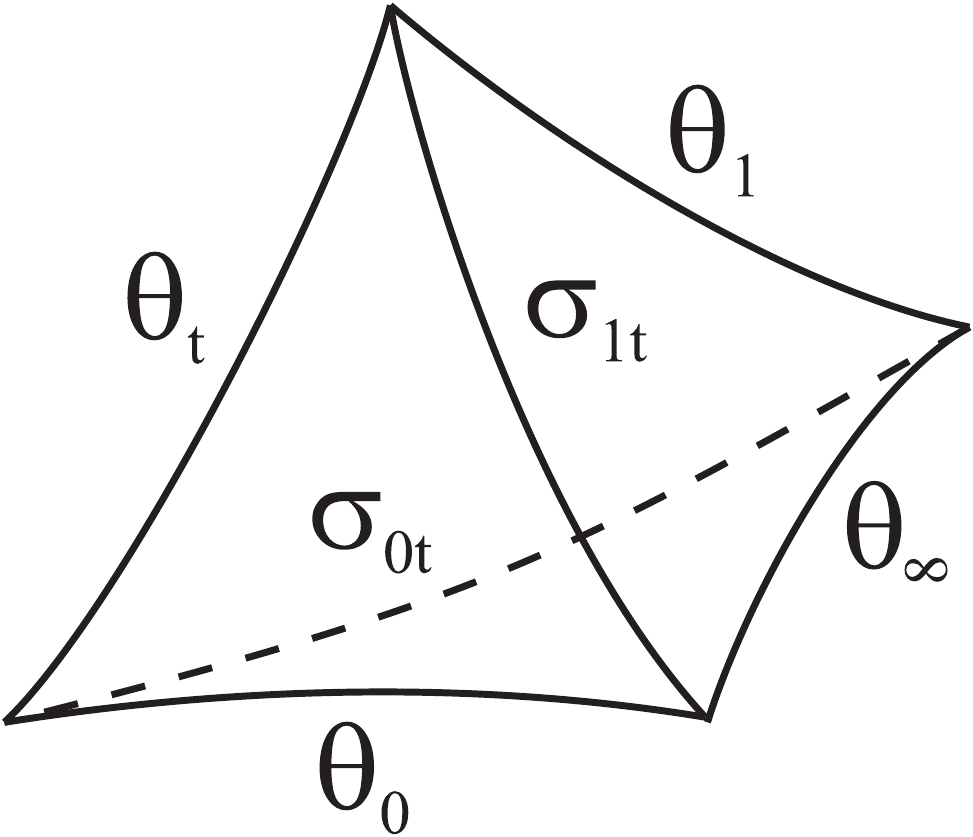}} \\
 {\small Fig. 2: Tetrahedron $\mathcal{T}$ obtained by gluing external legs of conformal blocks}
 \end{center}
 \end{figure}

 \subsection{The volume of $\mathcal{T}$}
 The explicit formula for the volume is most conveniently written in terms of the Lobachevsky function, which is essentially a half of the imaginary part of the Euler dilo\-garithm:
 \be\label{loba}
 \Lambda\left(\sigma\right)=\frac{1}{4i}\left[\mathrm{Li}_2\left(e^{2\pi i \sigma}\right)-\mathrm{Li}_2\left(e^{-2\pi i \sigma}\right)\right],\qquad \sigma\in\Rb.
 \eb
  This definition differs from the standard one by a factor of $\pi$ in the argu\-ment of~$\Lambda$.
  The dilogarithms are evaluated on their main sheets, which implies that
 $\Lambda(\sigma)$ is continuous and periodic.

 Define the parameters
 \begin{align}\label{nus}
 \begin{array}{ll}
 \nu_1=\sigma_{0t}+\theta_0+\theta_t, & \qquad\lambda_1=\theta_0+\theta_t+\theta_1+\theta_{\infty},\\
 \nu_2=\sigma_{0t}+\theta_1+\theta_{\infty},& \qquad\lambda_2=\sigma_{0t}+\sigma_{1t}+\theta_0+\theta_1, \\
 \nu_3=\sigma_{1t}+\theta_0+\theta_{\infty},& \qquad\lambda_3=\sigma_{0t}+\sigma_{1t}+\theta_t+\theta_{\infty},\\
 \nu_4=\sigma_{1t}+\theta_t+\theta_1,& \qquad\lambda_4=0,\\
 2\nu_{\Sigma}=\nu_1+\nu_2+\nu_3+\nu_4,&
 \end{array}
 \end{align}
 then the volume of $\mathcal{T}$ is given by \cite{Vol1,Murakami}
 \be\label{volt}
 \mathrm{Vol}\left(\mathcal{T}\right)=\frac{1}{2}\sum_{k=1}^4\sum_{\epsilon=\pm}\epsilon\left[
 \Lambda\left(\omega_{\epsilon}+\nu_k\right)-\Lambda\left(\omega_{\epsilon}+\lambda_k\right)\right].
 \eb
 Here $\omega_{\pm}$ represent two nontrivial solutions $z_{\pm}=e^{2\pi i \omega_{\pm}}$  of
 the equation
 \be\label{eqzpm}
 \prod_{k=1}^4\left(1-ze^{2\pi i \nu_k}\right)=\prod_{k=1}^4\left(1-ze^{2\pi i \lambda_k}\right).
 \eb
 which can be expressed in terms of $\vec{\theta}$, $\vec{\sigma}$ as
  \be\label{zpm}
   z_{\pm}=\frac{4\sin 2\pi\sigma_{0t}\sin2\pi\sigma_{1t}+4\sin 2\pi\theta_t\sin2\pi\theta_{\infty}+
 4\sin 2\pi\theta_0\sin2\pi\theta_{1}\pm q_{01}
 }{2\sum_{k=1}^4\left(e^{2\pi i \left(\nu_{\Sigma}-\nu_k\right)}-e^{2\pi i \left(\nu_{\Sigma}-\lambda_k\right)}\right)}.
 \eb
 Their product does not contain $p_{01}$ and can be written as
 \be\label{zpzm}
 z_+z_-=\frac{\sum_{k=1}^4\left(e^{2\pi i \left(\nu_k-\nu_{\Sigma}\right)}-e^{2\pi i \left(\lambda_k-\nu_{\Sigma}\right)}\right)}{\sum_{k=1}^4\left(e^{2\pi i \left(\nu_{\Sigma}-\nu_k\right)}-e^{2\pi i \left(\nu_{\Sigma}-\lambda_k\right)}\right)}.
 \eb

 Note that for the genuine hyperbolic tetrahedra ${\theta}$'s, $\sigma_{0t}$ and $\sigma_{1t}$ are real. Also, $q_{01}$ is purely imaginary
 since $\mathcal{G}$ should have the signature $\left(-,+,+,+\right)$ of the ambient space $\Rb^{1,3}\supset\mathbb{H}^3$. The para\-meters $z_{\pm}$ then lie on the
 unit circle, which makes (\ref{volt}) compatible with our earlier conventions for $\Lambda(\sigma)$.

 The precise relation between $\mathrm{Vol}\left(\mathcal{T}\right)$ and the generating function $\mathcal{S}(\vec{\theta};\sigma_{0t},\sigma_{1t})$ from the previous subsection is provided by
 \begin{lem}\label{lemma1}
 We have
 \begin{align}
 \label{der0t}
  2\frac{\partial\;}{\partial{\sigma_{0t}}}\mathrm{Vol}\left(\mathcal{T}\right)&\,=
 i\eta_{0t}+\pi\ln\prod_{\epsilon=\pm}\frac{\sin\pi\left(\theta_t+\sigma_{0t}+\epsilon\theta_0\right)
 \sin\pi\left(\theta_1+\sigma_{0t}+\epsilon\theta_{\infty}\right)}{
 \sin\pi\left(\theta_t-\sigma_{0t}+\epsilon\theta_0\right)
 \sin\pi\left(\theta_1-\sigma_{0t}+\epsilon\theta_{\infty}\right)},\\
 \label{der1t}
  -2\frac{\partial\;}{\partial{\sigma_{1t}}}\mathrm{Vol}\left(\mathcal{T}\right)&\,=i\eta_{1t}+\pi
 \ln\prod_{\epsilon=\pm}\frac{\sin\pi\left(\theta_t+\sigma_{1t}+\epsilon\theta_1\right)
 \sin\pi\left(\theta_0+\sigma_{1t}+\epsilon\theta_{\infty}\right)}{
 \sin\pi\left(\theta_t-\sigma_{1t}+\epsilon\theta_1\right)
 \sin\pi\left(\theta_0-\sigma_{1t}+\epsilon\theta_{\infty}\right)}.
 \end{align}
 \end{lem}

 Let us also mention that the formula (\ref{volt}) for  $\mathrm{Vol}\left(\mathcal{T}\right)$ can be rewritten
 in terms of $\omega_+$ or $\omega_-$  only (instead of using both of these parameters) thanks to the following result, cf \cite[Theorem~2]{Murakami}:
 \begin{lem}\label{lemma2}
 The quantity
 \begin{align}\label{halfsum}
 \mathcal{V}\left(\mathcal{T}\right)=\frac{1}{2}\sum_{k=1}^4\sum_{\epsilon=\pm}\left[
 \Lambda\left(\omega_{\epsilon}+\nu_k\right)-\Lambda\left(\omega_{\epsilon}+\lambda_k\right)\right],
 \end{align}
 can be alternatively expressed as
 \begin{align}
 \nonumber
  \mathcal{V}\left(\mathcal{T}\right)=\frac12\sum_{\epsilon,\epsilon'=\pm}\epsilon\epsilon'&\Bigl[
 \Lambda\left(\theta_0+\epsilon\sigma_{0t}+\epsilon'\theta_t\right)+
 \Lambda\left(\theta_{\infty}+\epsilon\sigma_{0t}+\epsilon'\theta_1\right)
 +\Bigl.\\
 \label{halfsum2}
 &\Bigl.+\Lambda\left(\theta_t+\epsilon\sigma_{1t}+\epsilon'\theta_1\right)+
 \Lambda\left(\theta_{\infty}+\epsilon\sigma_{1t}+\epsilon'\theta_0\right)\Bigr].
 \end{align}
 \end{lem}

 In the case of complex angles the function $\mathrm{Vol}\left(\mathcal{T}\right)$ may be defined via continuation
 from an open set $U\subset\Cb^6$. However, in doing this the periodicity with respect to angles will be lost,
 just as if instead
 of taking $\sigma\in\Rb$ and fixing the principal branches of $\mathrm{Li}_2(z)$ in (\ref{loba}) we tried to continue $\Lambda(\sigma)$ analytically from a suitable open subset of $\Cb$.

 \section{Connection coefficient for Painlev\'e VI tau function}\label{seccc}
 \subsection{Functional relations}
 In this section, we compute the connection coefficient defined by (\ref{concof}).
 The idea is to consider Painlev\'e VI parameters $\vec{\theta}$ as fixed and obtain $\chi_{01}(\vec{\theta};\sigma_{0t},\sigma_{1t};p_{01})$ by solving certain difference equations with respect to
 $\sigma_{0t}$ and $\sigma_{1t}$.

 Given $\vec{\theta}$, $\sigma_{0t}$ and $\sigma_{1t}$,  the value of $p_{01}$ which enters the tau function expansions (\ref{exp0})--(\ref{exp1}) via $s_{0t}$ and $s_{1t}$, is fixed up to the choice of solution of the
  Jimbo-Fricke equation $W\left(p_{0t},p_{1t},p_{01}\right)=0$.
 Therefore the space $\mathcal{M}_{\vec{\theta}}$ of triples $\left(\sigma_{0t},\sigma_{1t};p_{01}\right)$ associated to $\vec{\sigma}$
 at fixed $\vec{\theta}$ is a double cover of $\Cb^2\ni\left(\sigma_{0t},\sigma_{1t}\right)$. It may be safely assumed that
 $\chi_{01}(\vec{\theta};\sigma_{0t},\sigma_{1t};p_{01})$ is meromorphic on the complement of the ramification locus of $\mathcal{M}_{\vec{\theta}}$. The structure of conformal expansions (\ref{exp0})--(\ref{exp1}) then gives two recurrence relations for $\chi_{01}$:
 \begin{align}
 \label{frel1}
 \frac{\chi_{01}(\vec{\theta};\sigma_{0t}+1,\sigma_{1t};p_{01})}{
 \chi_{01}(\vec{\theta};\sigma_{0t},\sigma_{1t};p_{01})}=s_{0t}^{-1},\\
 \label{frel2}
  \frac{\chi_{01}(\vec{\theta};\sigma_{0t},\sigma_{1t}+1;p_{01})}{
 \chi_{01}(\vec{\theta};\sigma_{0t},\sigma_{1t};p_{01})}=s_{1t},
 \end{align}
 where $s_{0t}$, $s_{1t}$ are defined by (\ref{s0t})--(\ref{s1t}). The main difficulty in the solution of
 (\ref{frel1})--(\ref{frel2})
 is hidden in the dependence of these quantities on $p_{01}$, as the latter depends on $\sigma_{0t},\sigma_{1t}$ in a rather complicated way.

 As a warm-up exercise, let us consider the implications of the difference equations for the
 symmetrized product of connection
 coefficients over two sheets of $\mathcal{M}_{\vec{\theta}\,}$:
 \be\label{schi}
 \kappa_{01}(\vec{\theta};\sigma_{0t},\sigma_{1t})=
 \chi_{01}(\vec{\theta};\sigma_{0t},\sigma_{1t};p_{01})\chi_{01}(\vec{\theta};\sigma_{0t},\sigma_{1t};p_{01}').
 \eb
 Our notation means the following: if $p_{01}$ is one of the two roots of the equation $W(p_{0t},p_{1t},p_{01})=0$, then $p_{01}'=\omega_{0t}-p_{01}-p_{0t}p_{1t}$ denotes the other root. Also, denote by $q_{\mu\nu}'$, $s_{\mu\nu}'$ the appropriate modifications of $q_{\mu\nu}$, $s_{\mu\nu}$. In particular, one has
 \ben
 q_{0t}'=q_{0t}-p_{1t}q_{01},\qquad q_{1t}'=q_{1t}-p_{0t}q_{01},\qquad q_{01}'=-q_{01}.
 \ebn

 Now write $s_{0t}$, $s_{1t}$, $s_{0t}'$, $s_{1t}'$ in terms of $q_{01}$.
 It turns out, expectedly, that the products $s_{0t}s_{0t}'$ and $s_{1t}s_{1t}'$ depend only on $q_{01}^2$.
 Using (\ref{JFr2})
 and simplifying the corresponding expressions with the help of (\ref{minors1})--(\ref{minorsf}), it can be deduced from (\ref{frel1})--(\ref{frel2}) that
 \begin{align}
 \label{frel3}
 \frac{\kappa_{01}(\vec{\theta};\sigma_{0t}+1,\sigma_{1t})}{\kappa_{01}(\vec{\theta};\sigma_{0t},\sigma_{1t})}=
 \prod_{\epsilon=\pm}\frac{\sin\pi\left(\theta_t-\sigma_{0t}+\epsilon\theta_0\right)
 \sin\pi\left(\theta_1-\sigma_{0t}+\epsilon\theta_{\infty}\right)
 }{
 \sin\pi\left(\theta_t+\sigma_{0t}+\epsilon\theta_0\right)
 \sin\pi\left(\theta_1+\sigma_{0t}+\epsilon\theta_{\infty}\right)},\\
 \label{frel4}
 \frac{\kappa_{01}(\vec{\theta};\sigma_{0t},\sigma_{1t}+1)}{\kappa_{01}(\vec{\theta};\sigma_{0t},\sigma_{1t})}=
 \prod_{\epsilon=\pm}\frac{
  \sin\pi\left(\theta_t+\sigma_{1t}+\epsilon\theta_1\right)
 \sin\pi\left(\theta_0+\sigma_{1t}+\epsilon\theta_{\infty}\right)}{
 \sin\pi\left(\theta_t-\sigma_{1t}+\epsilon\theta_1\right)
 \sin\pi\left(\theta_0-\sigma_{1t}+\epsilon\theta_{\infty}\right)}.
 \end{align}

 The general solution of (\ref{frel3})--(\ref{frel4}) may be constructed in terms of Barnes $G$-function
 already encountered in Subsection~\ref{ssc25}.
 Indeed, since $G\left(1+\sigma\right)=\Gamma\left(\sigma\right)G\left(\sigma\right)$, the function
 \be\label{Ba}
 \hat{G}\left(\sigma\right)=\frac{G\left(1+\sigma\right)}{G\left(1-\sigma\right)}
 \eb
 satisfies
 \be\label{rechb}
  \hat{G}\left(\sigma\pm1\right)=\mp\left(\frac{\sin\pi\sigma}{\pi}\right)^{\mp1}\hat{G}\left(\sigma\right).
 \eb
 One then easily derives
 \begin{lem}
 The general solution of (\ref{frel3})--(\ref{frel4}) is given by
 \be\label{schilem}
 \kappa_{01}(\vec{\theta};\sigma_{0t},\sigma_{1t})=\kappa_{\mathrm{per}}(\vec{\theta};\sigma_{0t},\sigma_{1t})
 \prod_{\epsilon,\epsilon'=\pm}\hat{G}\biggl[\begin{array}{c}
 \theta_t+\epsilon\theta_0+\epsilon'\sigma_{0t},\theta_1+\epsilon\theta_{\infty}+\epsilon'\sigma_{0t}
 \\
 \theta_t+\epsilon\theta_1+\epsilon'\sigma_{1t},\theta_0+\epsilon\theta_{\infty}+\epsilon'\sigma_{1t}
 \end{array}\biggr],
 \eb
 where $\kappa_{\mathrm{per}}(\vec{\theta};\sigma_{0t},\sigma_{1t})$ is an arbitrary periodic function of both
 $\sigma_{0t},\sigma_{1t}$ with periods~1.
 \end{lem}

 What about $\kappa_{\mathrm{per}}(\vec{\theta};\sigma_{0t},\sigma_{1t})$? The simplest guess is to assume
 that this quantity does not depend on $\sigma_{0t},\sigma_{1t}$ (some arguments in favor
  of this hypothesis will be given in the next subsection). The guess is readily
 confirmed by numerical experiments,
 but in fact the numerics reveals much more: $\kappa_{\mathrm{per}}(\vec{\theta};\sigma_{0t},\sigma_{1t})$ is simply equal to 1!

 The final formula can
 now be written as
 \begin{align}\label{ffk01}
 \bar{\chi}_{01}(\vec{\theta};\sigma_{0t},\sigma_{1t};p_{01})\bar{\chi}_{01}(\vec{\theta};\sigma_{0t},\sigma_{1t};p_{01}')&=
 \frac{
 \Phi\left(\theta_t,\theta_1,\sigma_{1t}\right)\Phi\left(\theta_0,\theta_{\infty},\sigma_{1t}\right)}{
 \Phi\left(\theta_0,\theta_t,\sigma_{0t}\right)\Phi\left(\theta_1,\theta_{\infty},\sigma_{0t}\right)
 },
 \end{align}
 where
 \ben
 \Phi\left(\theta,\theta',\theta''\right)=\frac{\prod_{\epsilon,\epsilon',\epsilon''=\pm}G\left(1+\epsilon\theta+
 \epsilon'\theta'+\epsilon''\theta''\right)}{\prod_{\epsilon=\pm}G\left(1+2\epsilon\theta\right)
 G\left(1+2\epsilon\theta'\right)G\left(1+2\epsilon\theta''\right)}.
 \ebn
 The right side of  (\ref{ffk01}) coincides with a ratio of three-point functions $\Phi$
 in the time-like $c=1$ Liouville theory \cite{Zamo_Li}. A conceptual explanation
  of this intriguing coincidence is yet to be found.

  \subsection{Minimal solution}
  Let us now come back to the relations (\ref{frel1})--(\ref{frel2}).
  As their solution is much easier to check than to guess, the reader interested only in the final result
  may jump directly to Lemma~\ref{funcsols}. What follows is an attempt to elucidate the origins of this Lemma.

  Taking the logarithmic derivatives of both sides of (\ref{frel1})--(\ref{frel2})
  and recalling the parameterization (\ref{etas}), we obtain
 \begin{align}
 \label{derfr1}\frac{\partial\;}{\partial\sigma_{0t}}\ln\frac{\chi_{01}(\sigma_{0t}+1,\sigma_{1t})}{\chi_{01}(\sigma_{0t},\sigma_{1t})}=
 -\frac{1}{2\pi i}\frac{\partial\eta_{0t}}{\partial\sigma_{0t}},\\
  \label{derfr2}\frac{\partial\;}{\partial\sigma_{1t}}\ln\frac{\chi_{01}(\sigma_{0t}+1,\sigma_{1t})}{\chi_{01}(\sigma_{0t},\sigma_{1t})}=-\frac{1}{2\pi i}\frac{\partial\eta_{0t}}{\partial\sigma_{1t}},\\
 \label{derfr3} \frac{\partial\;}{\partial\sigma_{0t}}\ln\frac{\chi_{01}(\sigma_{0t},\sigma_{1t}+1)}{\chi_{01}(\sigma_{0t},\sigma_{1t})}=\;\;\;\frac{1}{2\pi i}\frac{\partial\eta_{1t}}{\partial\sigma_{0t}}\\
  \label{derfr4} \frac{\partial\;}{\partial\sigma_{1t}}\ln\frac{\chi_{01}(\sigma_{0t},\sigma_{1t}+1)}{\chi_{01}(\sigma_{0t},\sigma_{1t})}=\;\;\;\frac{1}{2\pi i}\frac{\partial\eta_{1t}}{\partial\sigma_{1t}},
 \end{align}
 where all other arguments of $\chi_{01}$ are temporarily omitted to lighten the notation.

 A tentative solution  of (\ref{derfr1})--(\ref{derfr4}) can be written in the form
 \be\label{tentsol}
 \ln\tilde{\chi}_{01}(\sigma_{0t},\sigma_{1t})=\frac{1}{2\pi i}\int_{P}^{(\sigma_{0t},\sigma_{1t})}\!\!\!\!
 \sigma_{1t}d\eta_{1t}-\sigma_{0t}d\eta_{0t},
 \eb
 where the integral is calculated along a path starting at some fixed point $P$ on the (infinite-sheeted covering of) Jimbo-Fricke surface. Indeed, the
 integrand is a closed 1-form; its differential $d\sigma_{1t}\wedge d\eta_{1t}-d\sigma_{0t}\wedge d\eta_{0t}$ vanishes since
 the transformation $(\sigma_{0t},\eta_{0t})\rightarrow (\sigma_{1t},\eta_{1t})$ is canonical, see Subsection~\ref{seccancoords}.
 Hence the integral value depends only on the homotopy class of the path on the Jimbo-Fricke surface with excluded one-dimensional
 subspaces corresponding to the singularities of the integrand. In particular, restricting to $\sigma_{1t}=\mathrm{const}$, one obtains a complex curve
 with punctures at the poles of the integrand.
 For $\tilde{\chi}_{01}$ defined by (\ref{tentsol}) one has, e.g.,
 \be\label{tentsol2}
 \ln\frac{\tilde{\chi}_{01}(\sigma_{0t}+1,\sigma_{1t})}{\tilde{\chi}_{01}(\sigma_{0t},\sigma_{1t})}=\frac{1}{2\pi i}\int_{(\sigma_{0t},\sigma_{1t})}^{(\sigma_{0t}+1,\sigma_{1t})}\!\!\!\!
 \sigma_{1t}d\eta_{1t}-\sigma_{0t}d\eta_{0t}.
 \eb
 Differentiating the right side, it is straightforward to check that $\tilde{\chi}_{01}$ satisfies (\ref{derfr1})--(\ref{derfr2}).
 The other two relations are verified analogously.

 Thus we have shown that, up to an additive constant independent of $\sigma_{0t}$, $\sigma_{1t}$ but a priori depending on
 homotopy class of integration path, $\ln\tilde{\chi}_{01}$ satisfies the same functional relations as $\ln\chi_{01}$.
 On the other hand, $\chi_{01}$ is expected to be a single-valued function of  $\sigma_{0t}$, $\sigma_{1t}$.
 If it were possible to present it in the form (\ref{tentsol}), the integrals corresponding to different paths could
 only differ by integer multiples of ${2\pi i}$. This appears not to be the case: the residues (e.g. calculated
 at the poles of the integrand restricted to $\sigma_{1t}=\mathrm{const}$) are not integers.
 Therefore, one may try to use the freedom in the choice of the additive constant to correct the integrand
 by a closed 1-form with periodic coefficients which would ensure the necessary analytic properties.

 The form of the correction term can be guessed as follows. The integral in (\ref{tentsol}) is obviously related to
 the generating function
 \be\label{sbranch}
 \mathcal{S}(\vec{\theta};\sigma_{0t},\sigma_{1t})=\int_{P}^{(\sigma_{0t},\sigma_{1t})}\!\!\!\!
 \eta_{0t}d\sigma_{0t}-\eta_{1t}d\sigma_{1t}
 \eb
 of the canonical transformation $(\sigma_{0t},\eta_{0t})\rightarrow (\sigma_{1t},\eta_{1t})$. This function is in turn
 related (see Lemma~\ref{lemma1}) to the complexified volume (\ref{volt}) of hyperbolic tetrahedron, expressed in terms
 of Lobachevsky functions. Both $\mathcal{S}(\vec{\theta};\sigma_{0t},\sigma_{1t})$ and $\mathrm{Vol}(\mathcal{T})$
 are multivalued functions of $\sigma_{0t}$, $\sigma_{1t}$,
 and we take the point of view that this multivaluedness stems from the possibility to consider
 homotopically inequivalent paths in (\ref{sbranch}). Now using that
 \be\label{lobbarnes}
 \Lambda(\sigma)=-\pi\sigma\ln\frac{\sin\pi\sigma}{\pi}-\pi\ln\hat{G}(\sigma),
 \eb
 one can decompose (\ref{tentsol}) into a Barnes function piece which has the required analytic behaviour,
 and an elementary function piece responsible for multivaluedness. It is natural to assume that the latter contribution
 can be compensated by the correction 1-form mentioned above, so that the genuine single-valued
 connection coefficient $\chi_{01}$ comes from (\ref{volt}) and Lemmas~\ref{lemma1}--\ref{lemma2} by keeping only
 log-Barnes term of (\ref{lobbarnes}) in each $\Lambda(\sigma)$. This finally leads to

  \begin{lem}\label{funcsols}
  The general solution of (\ref{frel1})--(\ref{frel2}) is given by
  \begin{align}\label{ffss}
   &\bar{\chi}_{01}(\vec{\theta};\sigma_{0t},\sigma_{1t};p_{01})=\chi_{\mathrm{per}}(\vec{\theta};\sigma_{0t},\sigma_{1t})\times\\
 \nonumber &\times\!\!\!\prod_{\epsilon,\epsilon'=\pm}\!\!G\biggl[\!
  \begin{array}{ll}
  1+\epsilon\sigma_{1t}+\epsilon'\theta_t-\epsilon\epsilon'\theta_1,1+\epsilon\sigma_{1t}+\epsilon'\theta_0-\epsilon\epsilon'\theta_{\infty}
  \\
  1+\epsilon\sigma_{0t}+\epsilon'\theta_t+\epsilon\epsilon'\theta_0,1+\epsilon\sigma_{0t}+\epsilon'\theta_1+\epsilon\epsilon'\theta_{\infty}
  \end{array}\!\biggr]\!\prod_{\epsilon=\pm}\frac{G(1+2\epsilon\sigma_{0t})}{G(1+2\epsilon\sigma_{1t})}\prod_{k=1}^4\frac{\hat{G}(\omega_++\nu_k)}{
  \hat{G}(\omega_++\lambda_k)},
  \end{align}
  where $\nu_{1\ldots 4}$ and $\lambda_{1\ldots 4}$ are defined by (\ref{nus}), $\omega_{+}$ by (\ref{eqzpm})--(\ref{zpm}), and $\chi_{\mathrm{per}}(\vec{\theta};\sigma_{0t},\sigma_{1t})$ is
  an arbitrary periodic function of $\sigma_{0t}$, $\sigma_{1t}$ with periods $1$.
  \end{lem}
  \noindent $\square$ Direct verification based on the identities of type (\ref{aux05}) used in the proof of Lemma~\ref{lemma1} in the
  Appendix. Observe that the right side of (\ref{ffss}) is a periodic function of $\omega_+$
  thanks to (\ref{eqzpm}) and  (\ref{rechb}), which enables one to choose the solution of $z_+=e^{2\pi i \omega_+}$
  arbitrarily.\hfill $\blacksquare$\vspace{0.1cm}

  The periodic prefactor $\chi_{\mathrm{per}}(\vec{\theta};\sigma_{0t},\sigma_{1t})$ can be fixed from the following considerations.
  Let $\sigma_{0t}=\alpha$ be a point where the connection coefficient tends to infinity.
  Unless this singular behavior is compensated by the coefficients of
  (\ref{exp0}) and (\ref{exp1}), all terms in the $t=0$ tau function expansion vanish whereas the $t=1$ series produces a nontrivial
  solution to Painlev\'e~VI. This contradiction suggests that $\chi_{\mathrm{per}}(\vec{\theta};\sigma_{0t},\sigma_{1t})$ is a
  nowhere vanishing holomorphic function of both $\sigma_{0t}$ and $\sigma_{1t}$. Making an additional assumption of nice behavior
  at infinity, one concludes that $\chi_{\mathrm{per}}(\vec{\theta};\sigma_{0t},\sigma_{1t})$ is in fact independent of
  the last two parameters.

  The remaining dependence on $\vec{\theta}$ can be strongly constrained using Painlev\'e~VI solutions known in closed form and depending
  on continuous parameters. For instance, such solutions are known for an infinite number of affine hyperplanes in the $\vec{\theta}$-space. In CFT language, they correspond to conformal blocks involving degenerate fields or (by Regge-Okamoto symmetry (\ref{regge})) to conformal blocks of
  free-field exponentials with screening insertions. The simplest nontrivial example
  of this type is given by
  \be\label{tauric}
  \tau(t)=t^{2\theta_0\theta_1}(1-t)^{2\theta_t\theta_1}{}_2F_1\left(1-2\theta_{\infty},2\theta_t;2\theta_0+2\theta_t; t\right),
  \eb
  where $\vec{\theta}$ are subject to the constraint $\theta_0+\theta_t+\theta_1+\theta_{\infty}=1$ and
  $\vec{\sigma}=(\theta_0+\theta_t,\theta_1+\theta_t,\theta_0+\theta_1)$.
  More general formulas can be found in Subsection~4.3 of \cite{p35}.
  Some further examples come from the continuous alge\-braic families
  of Painlev\'e~VI transcendents living on affine $\vec{\theta}$-subspaces of dimensions $1$ and $2$.

  It turns out that the connection coefficients computed directly in these particular cases are reproduced by the simplest
  possible ansatz
  $\chi_{\mathrm{per}}(\vec{\theta};\sigma_{0t},\sigma_{1t})=1$. It is further supported by numerical computations with random values of $\vec{\theta}$, $\vec{\sigma}$  and analytic checks using exceptional algebraic Painlev\'e~VI solutions \cite{boalch,dubrovin,kitaev,apvi}, discussed in Subsection~\ref{secpvichecks}.
  This transforms Lemma~\ref{funcsols} into the following
  \begin{claim}
  Connection coefficient (\ref{concof}), (\ref{concof2}) for the generic Painlev\'e VI tau function has the following expression in terms of monodromy data:
  \begin{align}
  \label{concofans}
  &\bar{\chi}_{01}(\vec{\theta};\sigma_{0t},\sigma_{1t};p_{01})=\\
  \nonumber &=\!\!\!\prod_{\epsilon,\epsilon'=\pm}\!\!G\biggl[\!
  \begin{array}{ll}
  1+\epsilon\sigma_{1t}+\epsilon'\theta_t-\epsilon\epsilon'\theta_1,1+\epsilon\sigma_{1t}+\epsilon'\theta_0-\epsilon\epsilon'\theta_{\infty}
  \\
  1+\epsilon\sigma_{0t}+\epsilon'\theta_t+\epsilon\epsilon'\theta_0,1+\epsilon\sigma_{0t}+\epsilon'\theta_1+\epsilon\epsilon'\theta_{\infty}
  \end{array}\!\biggr]\!\prod_{\epsilon=\pm}\frac{G(1+2\epsilon\sigma_{0t})}{G(1+2\epsilon\sigma_{1t})}\prod_{k=1}^4\frac{\hat{G}(\omega_++\nu_k)}{
  \hat{G}(\omega_++\lambda_k)}.
  \end{align}
  \end{claim}
  It is worth noting that the formula (\ref{concofans}) possesses a non-obvious symmetry: its right side remains
  invariant upon sign reversal of any of the parameters $\theta_{0,t,1,\infty}$, $\sigma_{0t}$ and~$\sigma_{1t}$.

 \subsection{Algebraic checks}\label{secpvichecks}
 Algebraic solutions of Painlev\'e~VI provide an instructive way to test
 the general expression (\ref{concofans}) for the connection coefficient. For example, Painlev\'e~VI equation with parameters
 $\vec{\theta}=\left(\frac14,\frac14,\frac14,\frac38\right)$ admits the solution (obtained from the Solution~30 of \cite{apvi})
  \begin{align}
 \nonumber \tau(t(s))&=\frac{\left(s^4-1\right)^{-\frac18}\left(s^4+1\right)^{-\frac{5}{192}}\left(i+(1-i)s+s^2\right)}{s^{\frac{1}{32}}
 (1+2s-s^2)^{\frac{7}{24}}(s^2+2s-1)^{\frac{1}{24}}
 (1+6s^2+s^4)^{\frac{1}{6}}}\times\\
 &\quad\times\left[\frac{(s^2+i)(1-2is+s^2)}{(s^2-i)(1+2is+s^2)}\right]^{\frac18},\label{ttt}\\
 \;\;\;t(s)&=-\frac{\left(1+s^2\right)^2\left(1-6s^2+s^4\right)^3}{32s^2\left(1+s^4\right)^3}.\label{ts}
 \end{align}
 This solution has 16 branches. Let us choose one of them, e.g. corresponding to the interval $s\in(1,\sqrt{2}+1)$.
 It is straightforward to check that the latter interval is bijectively mapped by (\ref{ts}) to $t\in(0,1)$. In particular,
 \begin{align}
 t\left(s\rightarrow\sqrt{2}+1\right) &\,\sim \frac{32}{27}\left(10-7\sqrt{2}\right)\left(\sqrt{2}+1-s\right)^3,\label{tnear0}\\
 1-t\left(s\rightarrow 1\right)&\,\sim 8\left(s-1\right)^2.\label{tnear1}
 \end{align}
 Expanding the tau function (\ref{ttt}) near the endpoints and using (\ref{tnear0})--(\ref{tnear1}), one also finds that
  \begin{align}
 \label{tym0}
 \tau\left(t\rightarrow0\right)&\,= 2^{\frac{197}{576}}\cdot 3^{\frac{1}{64}}\cdot e^{-i\phi}\cdot t^{-\frac{7}{72}}\left[1+O\left(t^{\frac23}\right)\right],\\
 \label{tym1}
 \tau\left(t\rightarrow1\right)&\,=  2^{\frac{5}{64}}\cdot (1-t)^{-\frac{1}{16}}
 \left[1+\frac38e^{-\frac{i\pi}{4}}\left(1-t\right)^{\frac12}+O\left(1-t\right)\right],
 \end{align}
 where the phase $\phi$ is a non-rational multiple of $\pi$,  explicitly given by
 \be
 \phi=\frac{1}{8}\left(\pi-\arctan\frac{7}{4\sqrt{2}}\right).
 \eb

 The asymptotics (\ref{tym0})--(\ref{tym1}) corresponds to monodromy exponents $\vec{\sigma}=\left(\frac16,\frac14,\frac16\right)$.
 The connection coefficient $\bar{\chi}_{01}$ can be computed directly from these formulas and the relation (\ref{concof3}).
 To show that the answer obtained in this way coincides with our expression (\ref{concofans}), it suffices to demonstrate the following identity:
 \begin{align}\label{exc}
 G\Biggl[\begin{array}{c}\frac{4}{3},\frac{1}{4},\frac{5}{4},\frac{5}{4},\frac{5}{4},\frac{1}{8},\frac{9}{8},
 \frac{11}{8},\frac{11}{8}\vspace{0.1cm}\\
 \frac{1}{2},\frac{3}{2},\frac{5}{3},\frac{5}{6},\frac{5}{6},\frac{13}{24},\frac{17}{24},
 \frac{23}{24},\frac{43}{24}\end{array}\Biggr]
 \prod_{k=1}^4\frac{\hat{G}(\omega_++\nu_k)}{
  \hat{G}(\omega_++\lambda_k)}=2^{-\frac{19}{72}}\cdot 3^{-\frac{1}{64}} \cdot e^{i\phi},
 \end{align}
 where
 \begin{align}
 &\omega_+=\frac{5}{48}+\frac{1}{4\pi i }\,\ln\left(2-\sqrt{\frac32}+\frac{1}{\sqrt{2}}\right),\\
 &(\nu_1,\nu_2,\nu_3,\nu_4)\;=\left(\frac{2}{3},\frac{19}{24},\frac{7}{8},\frac{3}{4}\right),\\
 &(\lambda_1,\lambda_2,\lambda_3,\lambda_4)=\left(\frac{9}{8},\frac{11}{12},\frac{25}{24},0\right).
 \end{align}
 The identity (\ref{exc}) is readily confirmed numerically by comparison of the first 500 significant digits
 at both sides. We have done similar checks of (\ref{concofans}) for more than
 $50$ branches of about $20$ exceptional algebraic Painlev\'e VI solutions.

 \section{Fusion matrix at $c=1$}\label{secfusion}
 \subsection{Relation to connection coefficient}\label{connectiontofusion}
 It is clear from the form of Painlev\'e VI tau function expansions (\ref{exp0})--(\ref{exp1}) that the connection coefficient (\ref{concof})
 is a close relative of the fusion matrix (\ref{fusion}) for $c=1$ conformal blocks. Let us now try to spell out their relation more explicitly.

 First observe that
 \begin{align}
 \nonumber & \bar{\mathcal{F}}_{1}\Bigl[\begin{array}{ll}\theta_1 & \theta_t \\ \theta_{\infty} & \theta_0\end{array};\sigma_{0t}\Bigr]\left(t\right)=
 \oint_{C_{\Lambda}}
 \frac{ds_{0t}}{2\pi i s_{0t}}\,\frac{\chi_0^{-1}\tau(t)}{C\Bigl[\begin{array}{ll}\theta_1 & \theta_t \\ \theta_{\infty} & \theta_0\end{array};\sigma_{0t}\Bigr]}=\\
 \label{sumfus}
 =&\,\oint_{C_{\Lambda}}
 \frac{ds_{0t}}{2\pi i s_{0t}}\sum_{n\in\mathbb{Z}}\bar{\chi}_{01}(\vec{\theta};\sigma_{0t},\sigma_{1t}+n;p_{01})\,
 \bar{\mathcal{F}}_{1}\Bigl[\begin{array}{ll}\theta_0 & \theta_t \\ \theta_{\infty} & \theta_1\end{array};\sigma_{1t}+n\Bigr]\left(1-t\right),
 \end{align}
 where $C_{\Lambda}$ denotes the circle $|s_{0t}|=e^{-2\pi\Lambda}$ in the complex $s_{0t}$-plane and the last equality
  is obtained by combining (\ref{exp0})--(\ref{exp1}) with the functional relation (\ref{frel2}).

 The next step is to transform the integral over $s_{0t}$ into an integral over $\sigma_{1t}$.
 The latter was considered so far as a function
 of $\sigma_{0t}$, $s_{0t}$ implicitly determined by (\ref{s0t})  --- recall that $s_{0t}$ parameterizes the
 pairs $(p_{1t},p_{01})$ at fixed $p_{0t}$. It is not difficult to show that
 \be\label{s0ts}
 p_{1t}=\frac{\alpha_+(\vec{\theta},\sigma_{0t})s_{0t}+\alpha_-(\vec{\theta},\sigma_{0t})s_{0t}^{-1}+(p_{0t}\omega_{01}-2\omega_{1t})}{
 p_{0t}^2-4},
 \eb
 with $\alpha_{\pm}(\vec{\theta},\sigma_{0t})$ given by
 \ben
 \alpha_{\pm}(\vec{\theta},\sigma_{0t})=16\prod\limits_{\epsilon=\pm}
 \sin\pi\left(\theta_t\mp\sigma_{0t}+\epsilon\theta_0\right)
 \sin\pi\left(\theta_1\mp\sigma_{0t}+\epsilon\theta_{\infty}\right).
 \ebn
 Observe that, as $\Lambda\rightarrow\infty$, $p_{1t}$ becomes very large, which means that either
 (a)  $\mathrm{Im}\,\sigma_{1t}\sim \Lambda $ so that $e^{2\pi i\sigma_{1t}}\sim s_{0t}$ or
 (b)  $\mathrm{Im}\,\sigma_{1t}\sim -\Lambda $, in which case $e^{2\pi i\sigma_{1t}}\sim s_{0t}^{-1}$.

  Since $\Lambda$ may indeed be chosen sufficiently large,
 (\ref{s0ts}) implies that
 $\sigma_{1t}$-integration contour may be chosen as a segment $[\sigma_{1t}^*,\sigma_{1t}^*+1]$ with $\mathrm{Im}\,\sigma_{1t}^*$
 being sufficiently large for all singularities of the integrand in (\ref{sumfus}) to be located below the line $\mathbb{R}+i\Lambda$.
 The sum over $n$ in (\ref{sumfus}) then produces an integral over the whole line so that
 \begin{align}
 \label{fusionf}
 \bar{\mathcal{F}}_{1}\Bigl[\begin{array}{ll}\theta_1 & \theta_t \\ \theta_{\infty} & \theta_0\end{array};\sigma_{0t}\Bigr]\left(t\right)=
 \int_{\mathbb{R}+i\Lambda}\!\!\!
 F\Bigl[\begin{array}{ll}\theta_1 & \theta_t \\ \theta_{\infty} & \theta_0\end{array};\begin{array}{c}\sigma_{1t} \\ \sigma_{0t}\end{array}\Bigr]
 \bar{\mathcal{F}}_{1}\Bigl[\begin{array}{ll}\theta_0 & \theta_t \\ \theta_{\infty} & \theta_1\end{array};\sigma_{1t}\Bigr]\left(1-t\right)
 d\sigma_{1t},
 \end{align}
 with
 \be\label{fusionfinal}
 F\Bigl[\begin{array}{ll}\theta_1 & \theta_t \\ \theta_{\infty} & \theta_0\end{array};\begin{array}{c}\sigma_{1t} \\ \sigma_{0t}\end{array}\Bigr]=
 \mu(\vec{\theta},\vec{\sigma})\cdot\bar{\chi}_{01}(\vec{\theta};\sigma_{0t},\sigma_{1t};p_{01})
 \eb
 and $\displaystyle\mu(\vec{\theta},\vec{\sigma})=\frac{1}{2\pi i}\frac{\partial \ln s_{0t}}{\partial \sigma_{1t}}$. The root
 $p_{01}$ of the Jimbo-Fricke relation is chosen
 as to reproduce the asymptotics $s_{0t}\rightarrow 0$ as $\mathrm{Im}\,\sigma_{1t}\rightarrow +\infty$.

 Lemma~\ref{lemma1}
 indicates that the prefactor $\mu(\vec{\sigma},\vec{\theta})$ from the last relation can be rewritten in a symmetric form
 \be\label{muf1}
 \mu(\vec{\theta},\vec{\sigma})=\frac{i}{2\pi^2}\frac{\partial^2}{\partial\sigma_{0t}\partial\sigma_{1t}}\mathrm{Vol}(\mathcal{T}),
 \eb
 where $\mathrm{Vol}(\mathcal{T})$ denotes the tetrahedral volume of Section~\ref{secpvi}. One can also obtain an explicit trigonometric
 expression
 \be\label{muf2}
 \mu(\vec{\theta},\vec{\sigma})=-\frac{4\sin2\pi\sigma_{0t}\sin2\pi\sigma_{1t}}{\sqrt{\mathrm{det}\,\mathcal{G}}},
 \eb
 where $\mathcal{G}$ denotes the Gram matrix  defined by (\ref{JFr2}) and the branch of the square root is chosen so that
 $\sqrt{\mathrm{det}\,\mathcal{G}}=q_{01}$.

 The formula (\ref{fusionf}) is a $c=1$ analog of the fusion relation (\ref{fusion}). One apparent difference is that here
 it becomes more convenient to integrate over a complex contour in the momentum space rather than $\mathbb{R}^+$.
 The integral kernel $F\Bigl[\begin{array}{ll}\theta_1 & \theta_t \\ \theta_{\infty} & \theta_0\end{array};\begin{array}{c}\sigma_{1t} \\ \sigma_{0t}\end{array}\Bigr]$ is the fusion matrix for $c=1$ conformal blocks. We emphasize that, up to an elementary prefactor,
 it coincides with Painlev\'e~VI connection constant explicitly given by (\ref{concofans}).

  \subsection{Numerics}
 The expression (\ref{fusionfinal}) for the fusion kernel can in principle be checked numerically. Fix, for instance,
 \ben
 \left(\begin{array}{l}
 \theta_0 \\ \theta_t \\ \theta_1 \\ \theta_{\infty}\end{array}\right)=
 \left(\begin{array}{l}
 0.49+0.42i \\ 0.64-0.31i \\ 0.28-0.35 i \\ 0.19+0.47i\end{array}\right),\qquad
 \sigma_{0t}=0.26+0.45i.
 \ebn

 To specify the integration contour in (\ref{fusionf}), one needs to analyze the singularities
 of $\bar{\chi}_{01}$ with respect to $\sigma_{1t}$. The poles (and zeros) may only
 be located at $\epsilon\theta_0+\epsilon'\theta_{\infty}+\Zb$, $\epsilon\theta_t+\epsilon'\theta_{1}+\Zb$, $\Zb/2$
 with $\epsilon,\epsilon'=\pm$. One also has square root branch points corresponding to zeros of $q_{01}$; in our case
 they are given by $\pm(0.40-0.37i)+\Zb$, $\pm(0.11+0.49i)+\Zb$. This results into the singularity structure shown in Fig.~3.
 Hence in the integration contour $\Rb+i\Lambda$ on the right of (\ref{fusionf}) one may set $\Lambda=1$.
 \vspace{-0.1cm}
 \begin{center}
 \resizebox{10cm}{!}{
 \includegraphics{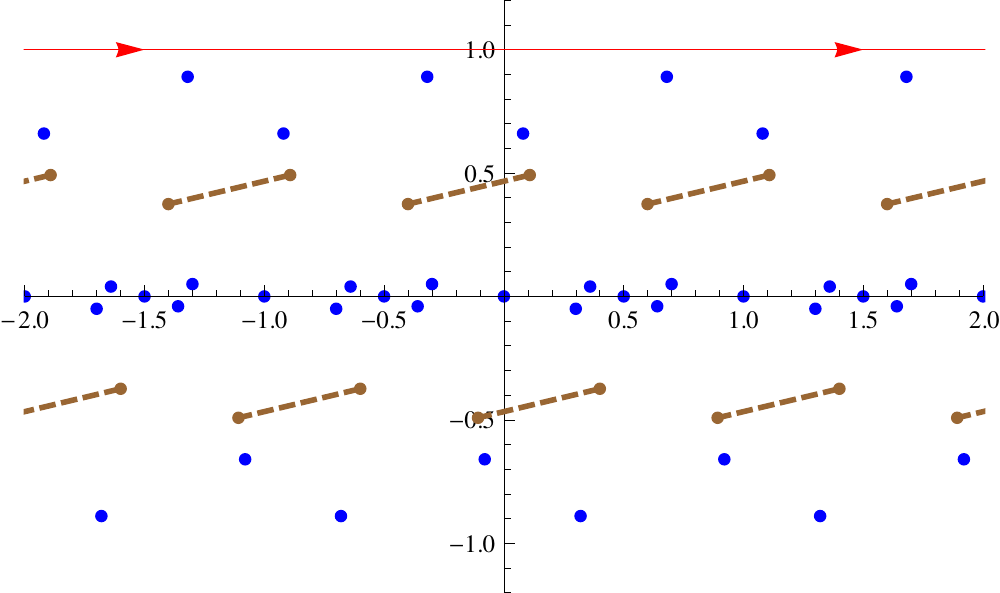}} \\ \vspace{-0.1cm}
 {\small Fig. 3: Singularities of $\bar{\chi}_{01}$ in the complex $\sigma_{1t}$-plane:
 the dots represent zeros and poles, the dashed lines correspond to branch cuts}
\end{center}\vspace{-0.1cm}

 In practice, the integrand decays very rapidly so that one can approximate the integral by the Riemann sum over uniform partition
 of the segment $[-1.5+i,2.5+i]$ into $400$ subintervals. Numerical values of conformal block
 functions were obtained
 by taking several first terms in their series expansions (30 on the left and 15 on the right of (\ref{fusionf})).

 Figure~4 shows the graphs of the left and right sides of (\ref{fusionf}),
 as functions of $t\in(0,1)$, calculated
 in this way.\vspace{-0.1cm}
\begin{center}
 \resizebox{14cm}{!}{
 \includegraphics{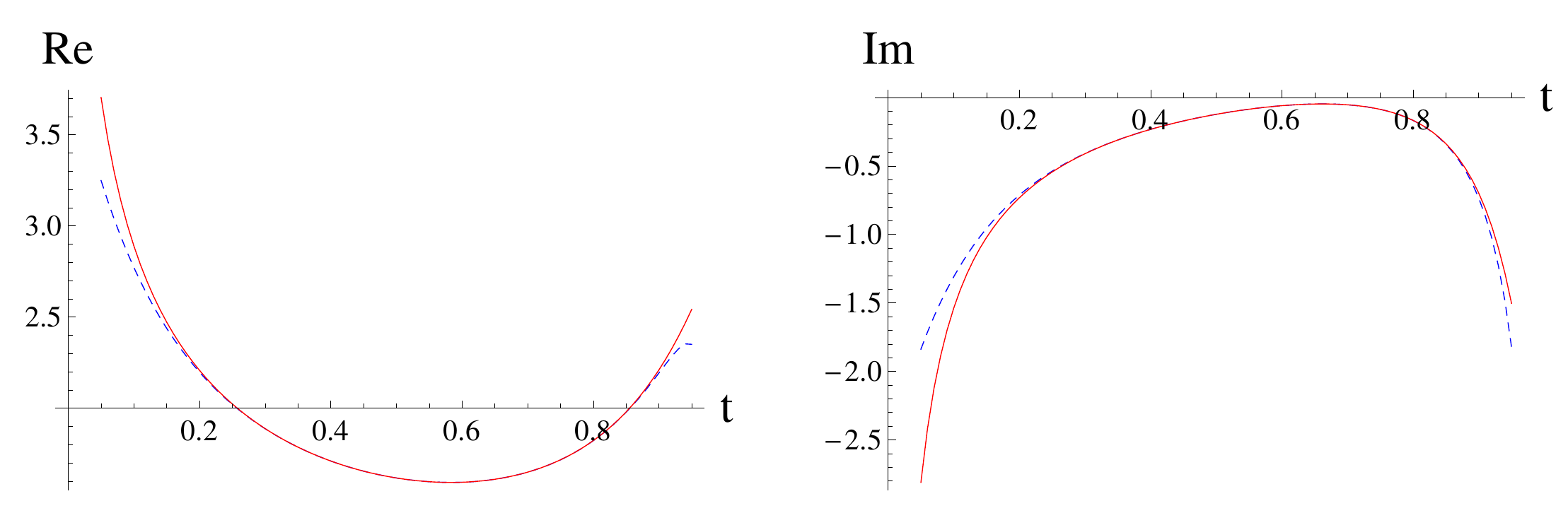}} \\ \vspace{-0.1cm}
 {\small Fig. 4: Real and imaginary parts of the l.h.s. (solid curve) and r.h.s. (dashed curve) of (\ref{fusionf})}
\end{center}
 Discrepancies at the endpoints are related to the fact that
 efficient appoximation of conformal block $\bar{\mathcal{F}}_1(t)$ by truncated series requires taking into account more and more
 terms  as $t$ approaches $1$.

 \subsection{Further checks: Ashkin-Teller conformal block}
 As $\theta_0=\theta_t=\theta_1=\theta_{\infty}=\frac14$, Jimbo-Fricke cubic (\ref{JFr}) reduces to
 \ben
 p_{0t}^2+p_{1t}^2+p_{01}^2+p_{0t}p_{1t}p_{01}=4.
 \ebn
 Considering $\sigma_{0t}$ and $\sigma_{1t}$ as fixed, one obtains the two possible solutions for $p_{01}$:
 \ben
 p_{01}=-2\cos2\pi(\sigma_{0t}\pm\sigma_{1t})
 \ebn
 which implies $q_{01}=\pm4\sin2\pi\sigma_{0t}\sin2\pi\sigma_{1t}$ and $s_{0t}=-e^{\mp2\pi i \sigma_{1t}}$. The parameter $\omega_+$ which
 appears in the con\-nec\-tion formula (\ref{concofans}) is then determined by
 \ben
 e^{2\pi i \omega_+}=-e^{-\pi i (\sigma_{0t}+\sigma_{1t})}\Bigl(\cos\pi(\sigma_{0t}-\sigma_{1t})\mp
 i \sin\pi(\sigma_{0t}+\sigma_{1t})\Bigr)^{\pm1}.
 \ebn
 Upon substitution of this expression into the connection formula (\ref{concofans}), the latter rather nontrivially simplifies to
 \ben
 \bar{\chi}_{01}(\sigma_{0t},\sigma_{1t};p_{01})=2^{4\sigma_{0t}^2-4\sigma_{1t}^2}e^{\pm 2\pi i \sigma_{0t}\sigma_{1t}}.
 \ebn
 The origin of this simplification is that here $\mu(\vec{\sigma})=\mp 1$, cf e.g. (\ref{muf1}) and (\ref{volt}).

 As explained in the previous subsections, the formulas (\ref{fusionf})--(\ref{fusionfinal}) come with a prescription for the choice of $p_{01}$.
 We should select from its two possible values the one characterized by vanishing of $s_{0t}$ as $\mathrm{Im}\,\sigma_{1t}\rightarrow\infty$. This corresponds to picking out the lower sign in the above formulas,
 and finally yields the folowing expression for the fusion matrix:
 \be\label{fusion_picard}
 F\Bigl[\begin{array}{ll}\frac14 & \frac14 \\ \frac14 & \frac14\end{array};\begin{array}{c}\sigma_{1t} \\ \sigma_{0t}\end{array}\Bigr]=
 2^{4\sigma_{0t}^2-4\sigma_{1t}^2}e^{- 2\pi i \sigma_{0t}\sigma_{1t}}.
 \eb
 Since the integral in (\ref{fusionf}) is calculated over the whole line, the last expression is
 clearly equivalent to the fusion kernel (\ref{fusionAT}) for Ashkin-Teller conformal blocks.

 \section{Discussion}
 We have used the recently established relation \cite{CFT_PVI,p35} of Painlev\'e equations and conformal field theory to
 solve two problems: the computation of the fusion matrix
 for generic $c=1$ conformal blocks and the connection problem for generic Painlev\'e~VI tau function.

 An important ingredient of the solution was the generating function of canonical transformations between two
 natural sets of Darboux coordinates on the Jimbo-Fricke cubic, related to the volume of hyperbolic tetrahedron.
 Our use of these objects was purely technical. However,  they seem to be a part of a bigger picture
 \cite{TV} relating isomonodromic
 deformations with both sides of AGT correspondence.
 The standard way to connect these theories is to study monodromy of $\mathfrak{sl}_2$-opers
 naturally appearing in CFT in the classical limit $c\to \infty$ \cite{Zamo_talk,NRS,TLang}.
 The next important step will be to achieve a proper understanding of the $c=1$ case.

  Conformal block (\ref{c1cb}) is also related to a large inter\-mediate
 dimension limit of the general conformal block. In \cite{Galakhov,Nemkov}, several first orders of perturbation theory in the
 so-called string coupling constant over this limiting point were calculated. Based on this calculation,
 it was conjectured that there are no perturbative corrections to the Fourier-type fusion kernel (\ref{fusionAT}).
 From this point of view,  the formulas (\ref{fusionfinal}), (\ref{concofans}) should include all nonperturbative
corrections in the string coupling constant. It would therefore be interesting to understand their relation to the results
 of \cite{Galakhov,Nemkov} in more detail.

  \section*{Appendix}
  \subsection*{Identities for minors of $\mathcal{G}$}
  Below we record several trigonometric identities satisfied by the first minors $M_{jk}\left(\mathcal{G}\right)$ of the Gram matrix $\mathcal{G}$
  defined by (\ref{JFr2}):
   \begin{align}
 \label{minors1}
 & M_{11}\left(\mathcal{G}\right)=-32\prod\nolimits_{\epsilon,\epsilon'=\pm}\sin\pi\left(\theta_1+\epsilon\theta_{\infty}+
 \epsilon'\sigma_{0t}\right),\\
 & M_{22}\left(\mathcal{G}\right)=-32\prod\nolimits_{\epsilon,\epsilon'=\pm}\sin\pi\left(\theta_t+\epsilon\theta_{1}+
 \epsilon'\sigma_{1t}\right),
 \\
 & M_{33}\left(\mathcal{G}\right)=-32\prod\nolimits_{\epsilon,\epsilon'=\pm}\sin\pi\left(\theta_0+\epsilon\theta_{\infty}+
 \epsilon'\sigma_{1t}\right),\\
 & M_{44}\left(\mathcal{G}\right)=-32\prod\nolimits_{\epsilon,\epsilon'=\pm}\sin\pi\left(\theta_t+\epsilon\theta_0+\epsilon'\sigma_{0t}\right),
 \\
 & M_{14}\left(\mathcal{G}\right)=2q_{1t}-p_{0t}q_{01},
 \\
 & M_{23}\left(\mathcal{G}\right)=2q_{0t}-p_{1t}q_{01},
 \\
 & M_{11}\left(\mathcal{G}\right)M_{44}\left(\mathcal{G}\right)=4\left(q_{1t}^2+q_{01}^2-p_{0t} q_{1t} q_{01}\right),\\
 \label{minorsf}
 & M_{22}\left(\mathcal{G}\right)M_{33}\left(\mathcal{G}\right)=4\left(q_{0t}^2+q_{01}^2-p_{1t} q_{0t} q_{01}\right).
 \end{align}
 All of these identities may be checked by direct calculation.

  \subsection*{Proof of Lemma~\ref{lemma1}}
   $\square$ For instance, let us compute the derivative of (\ref{volt}) with respect to $\sigma_{0t}$. Since
 $ \Lambda'\left(\sigma\right)=
 -\frac{\pi}{2}\ln\left(4\sin^2\pi \sigma\right)$
 and $z_{\pm}$ satisfy (\ref{eqzpm}), this derivative reduces to
 \begin{align}\label{aux01}
 \frac{\partial\;}{\partial{\sigma_{0t}}}\mathrm{Vol}\left(\mathcal{T}\right)=-\frac{\pi}{4}\ln
 \prod_{\epsilon=\pm}\left(\frac{\sin\pi\left(\omega_{\epsilon}+\nu_1\right)\sin\pi\left(\omega_{\epsilon}+\nu_2\right)}{
 \sin\pi\left(\omega_{\epsilon}+\lambda_2\right)
 \sin\pi\left(\omega_{\epsilon}+\lambda_3\right)}\right)^{2\epsilon}.
 \end{align}
 Using (\ref{zpm}), (\ref{zpzm}), (\ref{JFr2}) and some elementary algebra, it may be shown that
 \begin{align}
 \label{aux02}
 &\prod_{\epsilon=\pm}\frac{\sin\pi\left(\omega_{\epsilon}+\nu_1\right)\sin\pi\left(\omega_{\epsilon}+\nu_2\right)}{
 \sin\pi\left(\theta_t+\theta_0+\epsilon\sigma_{0t}\right)
 \sin\pi\left(\theta_1+\theta_{\infty}+\epsilon\sigma_{0t}\right)}=\\
 \label{aux03}
 =&\prod_{\epsilon=\pm}\frac{\sin\pi\left(\omega_{\epsilon}+\lambda_2\right)\sin\pi\left(\omega_{\epsilon}+\lambda_3\right)}{
 \sin\pi\left(\theta_t-\theta_0+\epsilon\sigma_{0t}\right)
 \sin\pi\left(\theta_1-\theta_{\infty}+\epsilon\sigma_{0t}\right)}=\\
 \label{aux04}
 =&\frac{16\prod_{\epsilon=\pm}\sin\pi\left(\theta_t-\theta_1+\epsilon\sigma_{1t}\right)
 \sin\pi\left(\theta_0-\theta_{\infty}+\epsilon\sigma_{1t}\right)}{\sum_{j,k=1}^4\left(e^{2\pi i \nu_j}-e^{2\pi i \lambda_j}\right)
 \left(e^{-2\pi i \nu_k}-e^{-2\pi i \lambda_k}\right)}.
 \end{align}
 Similarly, one has
 \begin{align}
 \nonumber
 \sin\pi\left(\omega_++\nu_1\right)\sin\pi\left(\omega_++\nu_2\right)
 \sin\pi\left(\omega_-+\lambda_2\right)\sin\pi\left(\omega_-+\lambda_3\right)=\\
 \label{aux05}
 =\frac{\prod_{\epsilon=\pm}
 \sin\pi\left(\theta_t-\theta_1+\epsilon\sigma_{1t}\right)
 \sin\pi\left(\theta_0-\theta_{\infty}+\epsilon\sigma_{1t}\right)}{
 \sum_{j,k=1}^4\left(e^{2\pi i \nu_j}-e^{2\pi i \lambda_j}\right)
 \left(e^{-2\pi i \nu_k}-e^{-2\pi i \lambda_k}\right)}\left(e^{2\pi i \sigma_{0t}}q_{01}-q_{1t}\right).
 \end{align}
 Decomposing the right side of (\ref{aux01}) into suitable combinations of (\ref{aux02}), (\ref{aux03}), (\ref{aux05}) and using the definition (\ref{s0t}) of $s_{0t}$, we obtain the equation (\ref{der0t}). The derivation of the second identity is
 completely analogous and will be omitted.
 \hfill$\blacksquare$

   \subsection*{Proof of Lemma~\ref{lemma2}}
 $\square$
 The only difference between the right sides of (\ref{halfsum}) and (\ref{volt}) is that the diffe\-rences
 of type $\Lambda\left(\omega_++\nu\right)-\Lambda\left(\omega_-+\nu\right)$ are replaced by the sums $\Lambda\left(\omega_++\nu\right)+\Lambda\left(\omega_-+\nu\right)$. Hence
 instead of (\ref{aux01}) one has
 \begin{align*}\label{aux01v2}
 \frac{\partial\mathcal{V}\left(\mathcal{T}\right)}{\partial{\sigma_{0t}}}=-\frac{\pi}{4}\ln
 \prod_{\epsilon=\pm}\left(\frac{\sin\pi\left(\omega_{\epsilon}+\nu_1\right)\sin\pi\left(\omega_{\epsilon}+\nu_2\right)}{
 \sin\pi\left(\omega_{\epsilon}+\lambda_2\right)
 \sin\pi\left(\omega_{\epsilon}+\lambda_3\right)}\right)^2.
 \end{align*}
 The ratio of (\ref{aux02}) and (\ref{aux03}) transforms this relation into
 \ben
 \frac{\partial\mathcal{V}\left(\mathcal{T}\right)}{\partial{\sigma_{0t}}}=
 -\frac{\pi}{4}\ln
 \prod_{\epsilon=\pm}\left(\frac{\sin\pi\left(\theta_t+\theta_0+\epsilon\sigma_{0t}\right)
 \sin\pi\left(\theta_1+\theta_{\infty}+\epsilon\sigma_{0t}\right)}{
 \sin\pi\left(\theta_t-\theta_0+\epsilon\sigma_{0t}\right)
 \sin\pi\left(\theta_1-\theta_{\infty}+\epsilon\sigma_{0t}\right)}\right)^2,
 \ebn
 and it is fairly easy to check that this coincides with the corresponding
 derivative of the right side of (\ref{halfsum2}). The derivative with respect to
 $\sigma_{1t}$ can be checked similarly.
 Now by continuity it suffices to verify (\ref{halfsum2}) for any fixed $(\sigma_{0t},\sigma_{1t})$. This can be
 done, for instance,
 for $\sigma_{0t}=-\theta_0-\theta_t$ and $\sigma_{1t}=\theta_1+\theta_t$, where we may set
 $\omega_+=0$, $\omega_-=-\lambda_1$.
 \hfill$\blacksquare$

 \end{document}